%% file: paper.tex
\documentclass[twocolumn, nofootinbib, amssymb]{revtex4-1}

\usepackage{amsmath}

\usepackage[usenames,dvipsnames]{color}

\usepackage[normalem]{ulem}

\usepackage{graphicx, hyperref}
\graphicspath{{Figures/}}

\newcommand{\half}{\tfrac{1}{2}}
\newcommand{\sixth}{\tfrac{1}{6}}
\newcommand{\tN}{\tilde N}
\def\diff{\mathrm{d}}
\def\ep{\varepsilon}
\DeclareMathOperator{\const}{const}

\begin{document}

\title{Critical phenomena in the general spherically symmetric
  Einstein-Yang-Mills system}

\author{Maciej Maliborski}
\email{maciej.maliborski@univie.ac.at}
\affiliation{Faculty of Physics, University of Vienna, Boltzmanngasse 5, A1090 Wien, Austria}
\affiliation{Max Planck Institute for Gravitational Physics
  (Albert Einstein Institute), Am M\"uhlenberg 1, 14476 Potsdam, Germany}

\author{Oliver Rinne}
\email{oliver.rinne@aei.mpg.de}
\affiliation{Hochschule f\"ur Technik und Wirtschaft Berlin,
  Treskowallee 8, 10318 Berlin, Germany}
\affiliation{Max Planck Institute for Gravitational Physics
  (Albert Einstein Institute), Am M\"uhlenberg 1, 14476 Potsdam, Germany}

\date{\today}

\begin{abstract}
  We study critical behavior in gravitational collapse of a general
  spherically symmetric Yang-Mills field coupled to the Einstein
  equations.  Unlike the magnetic ansatz used in previous numerical
  work, the general Yang-Mills connection has two degrees of freedom
  in spherical symmetry. This fact changes the phenomenology of
  critical collapse dramatically. The magnetic sector features both
  type I and type II critical collapse, with universal critical
  solutions. In contrast, in the general system type I disappears and
  the critical behavior at the threshold between dispersal and black
  hole formation is always type II. We obtain values of the mass
  scaling and echoing exponents close to those observed in the
  magnetic sector, however we find some indications that the critical
  solution differs from the purely magnetic discretely self-similar
  attractor and exact self-similarity and universality might be lost.
  The additional ``type III'' critical phenomenon in the magnetic
  sector, where black holes form on both sides of the threshold but
  the Yang-Mills potential is in different vacuum states and there is
  a mass gap, also disappears in the general system.  We support our
  dynamical numerical simulations with calculations in linear
  perturbation theory; for instance, we compute quasi-normal modes of
  the unstable attractor (the Bartnik-McKinnon soliton) in type I
  collapse in the magnetic sector.
\end{abstract}



\maketitle


\section{Introduction}
\label{s:intro}

The Einstein-Yang-Mills (EYM) equations form a particularly rich
dynamical system already in spherical symmetry. This is due to the
existence of nontrivial static and discretely self-similar solutions,
which play the role of unstable attractors.

The general spherically symmetric Yang-Mills (YM) connection has two
free potentials $w$ and $\omega$ (see Sec.~\ref{s:setup} for details).
Most numerical work so far
(e.g. \cite{Choptuik1996,Choptuik1999,Zenginoglu2008,Purrer2009,Bizon2010a,
  Rinne2014a}) has imposed in addition to spherical symmetry the
so-called \emph{magnetic ansatz} $\omega = 0$.  The term ``magnetic''
originates from the fact that for a static spacetime, the YM curvature
only has a magnetic part and no electric part in this case.  This
ansatz is self-consistent in the sense that if the initial data
satisfy $\omega=0$ then this remains so at all times.  In contrast, if
the so-called \emph{sphaleronic sector} is turned on by allowing
$\omega\neq 0$ in the initial data, then both $w$ and $\omega$ will be
nonzero during the evolution.  (The term ``sphaleron''
\cite{Lavrelashvili1995} appears to refer to similar solutions to the
Yang-Mills-Higgs equations; note there is no Higgs field here though.)
Hence the magnetic sector forms a subsystem of the most general
spherically symmetric EYM equations, which we sometimes also refer to
as the extended system.

As far as we know, so far the only numerical evolutions of the
extended system have been presented in \cite{Rinne2013}, even though
the equations have been worked out before, e.g.~in
\cite{Choptuik1999}. The paper \cite{Rinne2013} was mainly concerned
with power-law tails. The aim of the present paper is to study
critical phenomena in gravitational collapse in the extended system.

In critical collapse one chooses a one-parameter (usually denoted by
$p$) family of initial data such that (at least in the standard
definition) a black hole forms in the subsequent evolution for
$p > p_*$ and the field disperses to flat spacetime for $p < p_*$.
One now asks what happens close to the critical point $p=p_*$.  For
surveys of critical collapse of various matter models coupled to the
Einstein equations, we refer the reader to \cite{Bizon1996,
  Choptuik1999a, Gundlach2007}.

Let us first review the situation in the magnetic sector of the EYM
system.  Depending on the family of initial data, two different types
of critical behavior occur.

In type I critical collapse \cite{Choptuik1996}, black hole formation
for $p > p_*$ turns on at a finite (nonzero) value of the black hole
mass, and at the critical threshold the evolution approaches a static
solution.  This static solution is identified as the first member
$X_1$ of a discrete (countably infinite) family of regular static
solutions, the \emph{Bartnik-McKinnon solitons} \cite{Bartnik1988}.

In type II critical collapse \cite{Choptuik1996}, the black hole mass
$M$ vanishes as $p \searrow p_*$; more precisely,
$M\sim(p-p_*)^\gamma$ with an exponent $\gamma$ that is universal,
i.e.~independent of the particular family of initial data chosen. The
critical solution is discretely self-similar (for a definition see
Eq.~(\ref{e:scalingsymmetry}) below and \cite{Gundlach2007}). The
echoing exponent $\Delta$ related to the discrete self-similarity as
well as the critical solution itself are universal.

There is a third type of critical collapse, which unlike the other two
is specific to the YM field used here as a matter model.  This is
related to the fact that assuming spherical symmetry and the magnetic
ansatz, there are two values of the potential $w$, namely $w=\pm 1$,
that both correspond to vacuum (i.e.~vanishing YM curvature and hence
energy-momentum tensor). In type III collapse one considers a family
of initial data that lead to black hole formation for all values of
the parameter $p$, but such that the final value of the YM potential
is $w=1$ for $p>p_*$, say, and $w=-1$ for $p<p_*$. Even though both
outcomes correspond to a vacuum black hole, the dynamical evolutions
are different and the black hole mass is discontinuous across the
threshold \cite{Choptuik1999,Rinne2014a}. The critical solution is
static (as in type I) and is identified with the first member $Y_1$ of
a discrete family of static hairy (i.e. with nonzero YM field) black
hole solutions, the \emph{colored black holes}
\cite{Bizon1990,Volkov1990}.

For a static or self-similar solution to appear as a critical solution
in a one-parameter bisection search, this solution must have precisely
one unstable mode when considering linear perturbations
\cite{Gundlach2007}. Linear perturbations of the Bartnik-McKinnon
solitons $X_n$ and colored black holes $Y_n$ were studied in
\cite{Lavrelashvili1995,Volkov1995}. In the magnetic sector $X_n$ and
$Y_n$ both have $n$ unstable modes. So indeed $X_1$ and $Y_1$ have
precisely one unstable mode in the magnetic sector. However, in the
extended system $X_n$ and $Y_n$ have a total of $2n$ unstable modes.
Thus $X_1$ and $Y_1$ now have two unstable modes, and hence they
cannot be codimension-one unstable attractors in the extended system.
This indicates that the phenomenology of critical collapse is likely
to be very different. It is important to note here that subject to
suitable falloff conditions, there are no static solutions with
nonzero electric part of the YM curvature except for the
Reissner-Nordstr\"om solution \cite{Galtsov1989,Ershov1990}. Hence no
nontrivial potential static attractors are added when moving from the
magnetic to the general ansatz.

One of our main results is that there is no type I critical collapse
in the extended system, instead the critical behavior at the threshold
between dispersal and black hole formation is always type II. We
compare the critical solution and scaling exponents with those in the
magnetic sector. For small sphaleronic perturbations the
Bartnik-McKinnon soliton $X_1$ can be observed as an intermediate
attractor before the self-similar type II critical solution is
approached. We study in detail how the type II mass scaling sets on
when perturbing off data that in the magnetic sector would be type
I-critical.

We also refine some results in the magnetic sector, namely we find
wiggles on top of the power-law scaling of the curvature in
subcritical evolutions, which allow for an independent estimate of the
type II echoing exponent. In type I collapse in the magnetic sector,
we show how $X_1$ is approached via a quasinormal mode (QNM) and a
tail, and we compare with a calculation of the QNM frequency in linear
perturbation theory.

Concerning type III collapse, once a small sphaleronic perturbation in
$\omega$ is added, the discontinuous transitions in the YM potential
$w$ and the black hole mass $M$ across the critical threshold are
replaced by continuous ones. Thus there is no critical behavior any
longer. In the magnetic sector we find tentative evidence of a QNM
ringdown to the colored black hole critical solution.

Our numerical results were obtained with two independent codes using
different coordinates. The type I and type II simulations employ
standard polar-areal (Schwarzschild-like) coordinates. For type III
collapse we use hyperboloidal slices of constant mean curvature, which
are conformally compactified towards future null infinity. The details
of and motivations for these different coordinate choices are
explained in Sec.~\ref{s:setup}.

This paper is organized as follows. In Sec.~\ref{s:setup} we describe
our ansatz for YM connection in spherical symmetry and our choices of
spacetime coordinates. Our numerical results on type I and type II
critical collapse are presented in Sec.~\ref{s:typeI_II}, and on type
III collapse in Sec.~\ref{s:typeIII}. We conclude in
Sec.~\ref{s:concl}. Further details are deferred to the appendices:
the equations solved by our two codes are given in Appendix
\ref{s:fieldeqns}, linear perturbations of the static solutions are
analyzed in Appendix \ref{s:linpert}, and a brief summary of our
numerical methods can be found in Appendix \ref{s:nummethods}.


\section{Setup and coordinate choices}
\label{s:setup}

The most general spherically symmetric YM connection with gauge group
SU(2) can be written in the following form after exploiting the
residual SU(2) gauge freedom \cite{Witten1977,Gu1981}:
\begin{multline}
  \label{e:ymansatz}
  \mathcal{A} = u\tau_{3}\diff t
  + \left(w\tau_{1}+\omega\tau_{2}\right)\diff\theta
  \\
  + \left(\cot{\theta}\tau_{3}+w\tau_{2}-\omega\tau_{1}\right)
  \sin{\theta}\diff\phi,
\end{multline}
where $u$, $w$ and $\omega$ are functions of $t$ and $r$ only and
$\tau_{i}$ form a standard basis of SU(2),
$\left[\tau_{i},\tau_{j}\right] = \varepsilon_{ijk}\tau_{k}$, where
$\varepsilon_{ijk}$ is totally antisymmetric with
$\varepsilon_{123}=1$.

An alternative parametrization of the YM connection, used in
\cite{Rinne2013}, is
\begin{align}
  \label{e:ymansatz2}
  \mathcal{A}^{i(a)} &=
  \varepsilon^{aij} x^j F + (x^a x^i - r^2 \delta^{ai}) H,
  \nonumber \\ \mathcal{A}_0^{(a)} &= G x^a,
\end{align}
where $(a)$ denotes the SU(2) gauge group index, all indices run over
$1,2,3$ and repeated indices are summed over. The field equations that
these two parametrizations give rise to are equivalent; the
correspondence between the variables is\footnote{The gauge
  transformation
  $\mathcal{A} \rightarrow U\mathcal{A}U^{-1} + U\diff{U^{-1}}$ with
  $U = e^{\theta \tau_{1}} e^{(\pi/2-\varphi)\tau_{3}}$ transforms
  (\ref{e:ymansatz2}) into (\ref{e:ymansatz}). We note that changing
  the sign of $w$ and $\omega$ simultaneously leaves the field
  equations invariant.}
\begin{equation}
  F = \frac{1+w}{r^{2}},\ H=-\frac{\omega}{r^{3}},\ G=\frac{u}{r}.
\end{equation}

The magnetic ansatz consists in setting $\omega=u=0$ (or equivalently
$H=G=0$). It leads to a self-consistent set of field equations. It
should be stressed that the additional YM potential $\omega$ (or
equivalently $H$) in the general ansatz \eqref{e:ymansatz}
\emph{cannot} be transformed away by an SU(2) gauge transformation; it
forms a second physical degree of freedom, the sphaleronic sector.
The function $u$ (or equivalently $G$) on the other hand can be
thought to be determined by $w$ and $\omega$ via the YM constraint
equation (cf.~Appendix \ref{s:fieldeqns}).

We have implemented two different choices of spacetime coordinates.
For the simulations of type I and type II critical collapse presented
below, we use polar-areal coordinates, in which the line element takes
the form
\begin{equation}
  \label{e:ds2_polar-areal}
  d s^{2} = -Ae^{-2\delta}\diff t^{2} + \frac{\diff r^{2}}{A}
  + r^{2} d\sigma^{2},
\end{equation}
where $d\sigma^{2}$ denotes the standard round metric on the
two-sphere.

For the simulations of type III critical collapse, we use
constant-mean-curvature (CMC) slices and isotropic spatial
coordinates,
\begin{equation}
  \label{e:ds2_CMC}
  d s^{2} = \Omega^{-2} [-\tilde N^2 \diff{t}^{2} +
  (\diff r + rX\diff t)^2 + r^2 d\sigma^2].
\end{equation}
The reason is that black holes form on both sides of the critical
threshold in type III collapse, and polar slices cannot penetrate
black hole horizons, whereas CMC slices can. Furthermore, CMC slices
extend to future null infinity, which provides a natural boundary of
the computational domain where no boundary conditions need to be
imposed as all the characteristics leave the domain. Hence very long
evolutions unspoilt by any effects of an artificial timelike outer
boundary are possible.

The EYM field equations in the two different formulations are given in
Appendix \ref{s:fieldeqns}.


\section{Type I and type II collapse}
\label{s:typeI_II}

In this section we present our numerical results on type I and type II
critical behavior both in the magnetic sector and the sphaleronic
sector. These simulations were carried out using the code based on
polar-areal coordinates.


\subsection{Initial data}

In our studies of critical phenomena we experimented with different
choices of initial data but for clarity we present our results for
three particular families:
\begin{itemize}
\item[(i)] a localized Gaussian perturbation
  \begin{align}
    \label{e:idw}
    w(0,r) &= 1 + a_{1}\exp\left[
             -\left(\frac{r-x_{1}}{s_{1}}\right)^{2\,q_{1}}\right],
    \\
    \label{e:idomega}
    \omega(0,r) &= a_{2} \left(\frac{r}{x_{2}}\right)^{3} \exp\left[
                  -\left(\frac{r-x_{2}}{s_{2}}\right)^{2\,q_{2}}\right],
    \\
    \label{e:idPi}
    \Pi(0,r) &= \partial_{r} w(0,r),
    \\
    \label{e:idP}
    P(0,r) &= 0,
  \end{align}
\item[(ii)] kinklike data
  \begin{align}
    \label{eq:idw2}
    w(0,r) &= 1 - a_{1}\tanh \left(\frac{r}{s_{1}}\right)^{q_{1}} ,
    \\
    \label{eq:idomega2}
    \omega(0,r) &= - a_{2}\tanh \left(\frac{r}{s_{2}}\right)^{q_{2}} ,
    \\
    \label{e:idPi2}
    \Pi(0,r) &= \frac{r}{s_{1}}\partial_{r} w(0,r),
    \\
    \label{e:idP2}
    P(0,r) &= \frac{r}{s_{2}}\partial_{r}\omega(0,r),
  \end{align}
\item[(iii)] and purely magnetic kinklike data
  \begin{align}
    \label{eq:idw3}
    w(0,r) &= 1 + a_{1}\left[-2 \tanh\left(\frac{r}{s_{1}}\right)^{q_{1}}
           + 2\tanh\left(\frac{r}{s_{2}}\right)^{q_{2}}\right],
    \\
    \label{eq:idomega3}
    \omega(0,r) &= 0,
    \\
    \label{e:idPi3}
    \Pi(0,r) &= a_{2} \left[\frac{r}{s_{1}}\partial_{r}
             \left(-2 \tanh\left(\frac{r}{s_{1}}\right)^{q_{1}}\right)
             \right.\\
             & \qquad \quad \left.+\frac{r}{s_{2}}\partial_{r}
             \left(2\tanh\left(\frac{r}{s_{2}}\right)^{q_{2}}\right)
             \right],
    \\
    \label{e:idP3}
    P(0,r) &= 0.
  \end{align}
\end{itemize}
Here the auxiliary variables $\Pi$ and $P$ are essentially time
derivatives of $w$ and $\omega$ [cf. Eqs. \eqref{e:dtw},
\eqref{e:dtomega}] and are set to make $w$ and $\omega$ either
approximately ingoing or stationary initially. We note that this
parametrization of the fields has been chosen to be consistent with
the following regularity conditions at the origin, which follow from a
Taylor expansion of the field equations:
\begin{equation}
  w = 1 + \mathcal{O}(r^2), \quad \omega = \mathcal{O}(r^3)
\end{equation}
(from which we also get analogous behavior of $\Pi$ and $P$, see
(\ref{e:dtw})-(\ref{e:dtomega})). Asymptotic flatness requires
\begin{equation}
  \label{eq:8}
  w^{2} + \omega^{2} \rightarrow 1 \text{ as } r\rightarrow \infty,
\end{equation}
which has to be satisfied by the initial data; in particular for
kink-like data (ii) this condition introduces the constraint
$a_{2}^{2} = a_{1}(2-a_{1})$. The choice of parameters will be
discussed below depending on the situation considered.


\subsection{Magnetic sector}
\label{s:typeI_II_mag}

We begin by restricting ourselves to the magnetic sector for family
(i), i.e. $a_2 = 0$ in \eqref{e:idomega}. Here we observe both type I
and type II critical behavior as previously analyzed in
\cite{Choptuik1996}.

\subsubsection{Type II collapse}
\label{sec:type-ii-collapse}

First we investigate type II critical collapse.  For this we vary
$p := a_1$ in \eqref{e:idw} and fix the remaining parameters to
$s_1 = 1/4$, $x_1 = 3$ and $q_1 = 1$. The value of the critical
amplitude is found to be $p_{*}\approx 0.14783$. We observe a
universal scaling of the mass of the apparent horizon in supercritical
evolutions
\begin{equation}
  \label{eq:15}
  M_\mathrm{AH} \sim (p-p_*)^\gamma
\end{equation}
with $\gamma = 0.20018\pm 0.00017$, and also a polynomial scaling of
$\mathcal{R}^{2} := \left. R_{\mu\nu} R^{\mu\nu}\right|_{r=0}$ in
subcritical evolutions
\begin{equation}
  \label{eq:9}
  \mathcal{R}^{2} \sim (p-p_*)^{-4\gamma}
\end{equation}
with the exponent $-4\gamma = -0.7886\pm 0.0029$, i.e.
$\gamma=0.19714\pm 0.00074$. These values for $\gamma$ are consistent
with the value $\gamma \approx 0.20$ reported in \cite{Choptuik1996}
and with the result $\gamma = 0.1964 \pm 0.0007$ obtained by directly
computing the critical solution and its perturbations
\cite{Gundlach1997a}.  The discrepancy of the super- and subcritical
scaling exponents $\gamma$ obtained from time evolutions of near
critical data results mainly from the inaccurate estimate of the
apparent horizon in the supercritical case. The scaling exponent we
find in subcritical evolutions is much more accurate and is closer to
the value of \cite{Gundlach1997a}.

\begin{figure}[!th]
  \centering
  \includegraphics[width=0.47\textwidth]{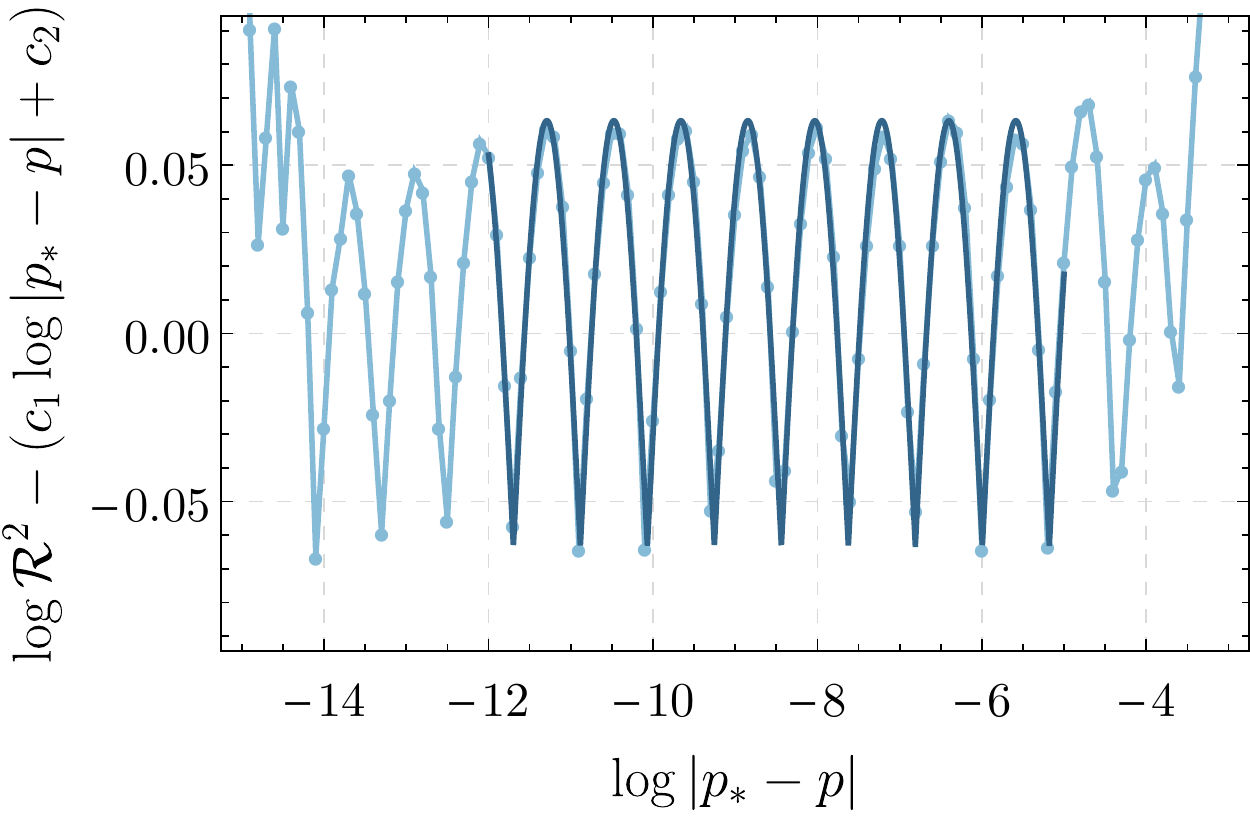}
  \caption{An analysis of subcritical data in type II critical
    behavior. Data (points) of $\log{\mathcal{R}^{2}}$ are compared
    with a five-parameter fit
    $c_{1} q + c_{2} + c_{3}\left|\cos(c_{4}q+c_{5})-1/2\right|$,
    where $q := \log\left|p_{*}-p\right|$, after subtraction of the
    linear part. Our choice of periodic function for $q$ is rather
    \textit{ad hoc} and is justified by its relatively good agreement
    with the numerical data.}
  \label{f:MagneticWiggles}
\end{figure}

In a graph of $\log{\mathcal{R}^{2}}$ vs.  $\log\left|p_{*}-p\right|$
we see periodic wiggles on top of the straight line, which are shown
in Fig.~\ref{f:MagneticWiggles}.  From the fit to the numerical data
we determine the period of oscillation to be
$\tau_{\mathcal{R}} \approx 0.815$, which is roughly comparable to the
theoretical prediction in \cite{Hod1997},
$\Delta/(4\gamma)\approx 0.939$ (with the values of $\Delta$ and
$\gamma$ taken from \cite{Gundlach1997a}), where $\Delta$ is the
echoing exponent discussed in the following.

\begin{figure}[!th]
  \centering
  \includegraphics[width=0.475\textwidth]{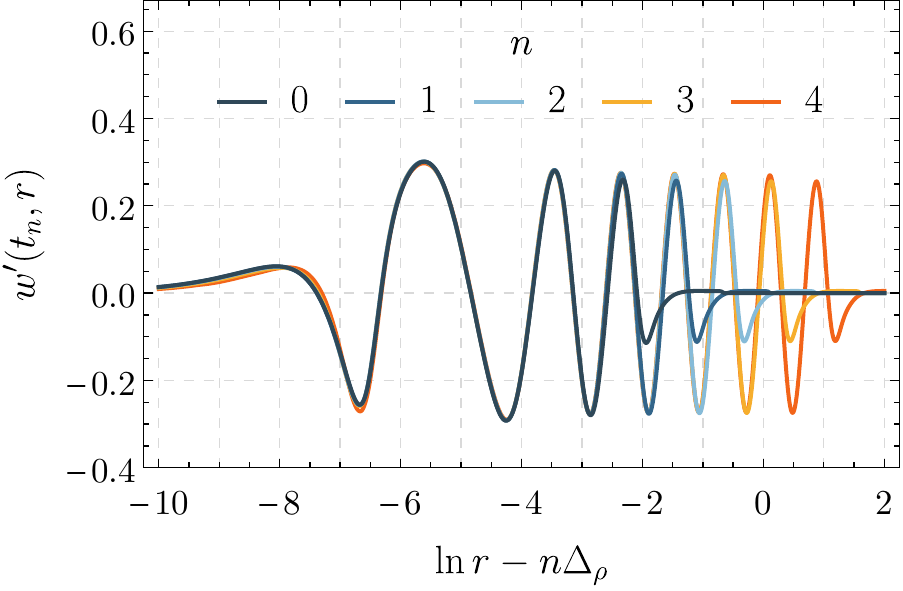}
  \caption{An illustration of the discrete self-similarity of the
    critical solution in type II critical collapse within the magnetic
    ansatz. This plot should be compared with Fig.~3 of
    \cite{Choptuik1996}.}
  \label{f:MagneticPeriodicity}
\end{figure}

The solution in the near-critical regime shows the approximate scaling
symmetry\footnote{That this is only an approximate symmetry
  follows from the existence of a scale in the EYM system set by the YM
  coupling constant. However close to the critical point this scale
  becomes irrelevant and thus (\ref{e:scalingsymmetry}) holds
  \cite{Gundlach1997a}.}
\begin{equation}
  \label{e:scalingsymmetry}
  Z(\tau - \Delta_\tau, \rho - \Delta_\rho) = Z(\tau, \rho)
\end{equation}
for a scale-free variable $Z$ in terms of logarithmic coordinates
\begin{equation}
  \label{eq:11}
  \rho = \ln r, \quad \tau = \ln (T_0^* - T_0),
\end{equation}
where $T_0$ denotes proper time at the origin\footnote{ $T_0$
  coincides with the coordinate time $t$ used in our code (see
  Appendix \ref{s:fieldeqns}).}  and $T_0^*$ is the accumulation time
of the type II critical solution. This is depicted for the scale-free
variable $w' := \partial_{r}w$ in Fig.~\ref{f:MagneticPeriodicity}.
The spatial echoing exponent $\Delta_\rho$ determined by rescaling the
spatial profiles at times at which the profiles overlap is found to be
$\Delta_{\rho}\approx 0.736\pm 0.001$.  From a discrete set of such
matching times we estimate the temporal period
$\Delta_{\tau}\approx 0.7364 \pm 0.0007$ (we also get an estimate for
the collapse time $T_{0}^{*}$, however this depends on the initial
data). These results support the claim that
$\Delta_\rho = \Delta_\tau =: \Delta$ and are consistent with the
value $\Delta \approx 0.74$ reported in \cite{Choptuik1996} as well as
the refined value $\Delta = 0.73784 \pm 0.00002$ in
\cite{Gundlach1997a}.

\begin{figure*}[!ht]
  \centering
  \begin{tabular}{lr}
    \includegraphics[width=0.475\textwidth]{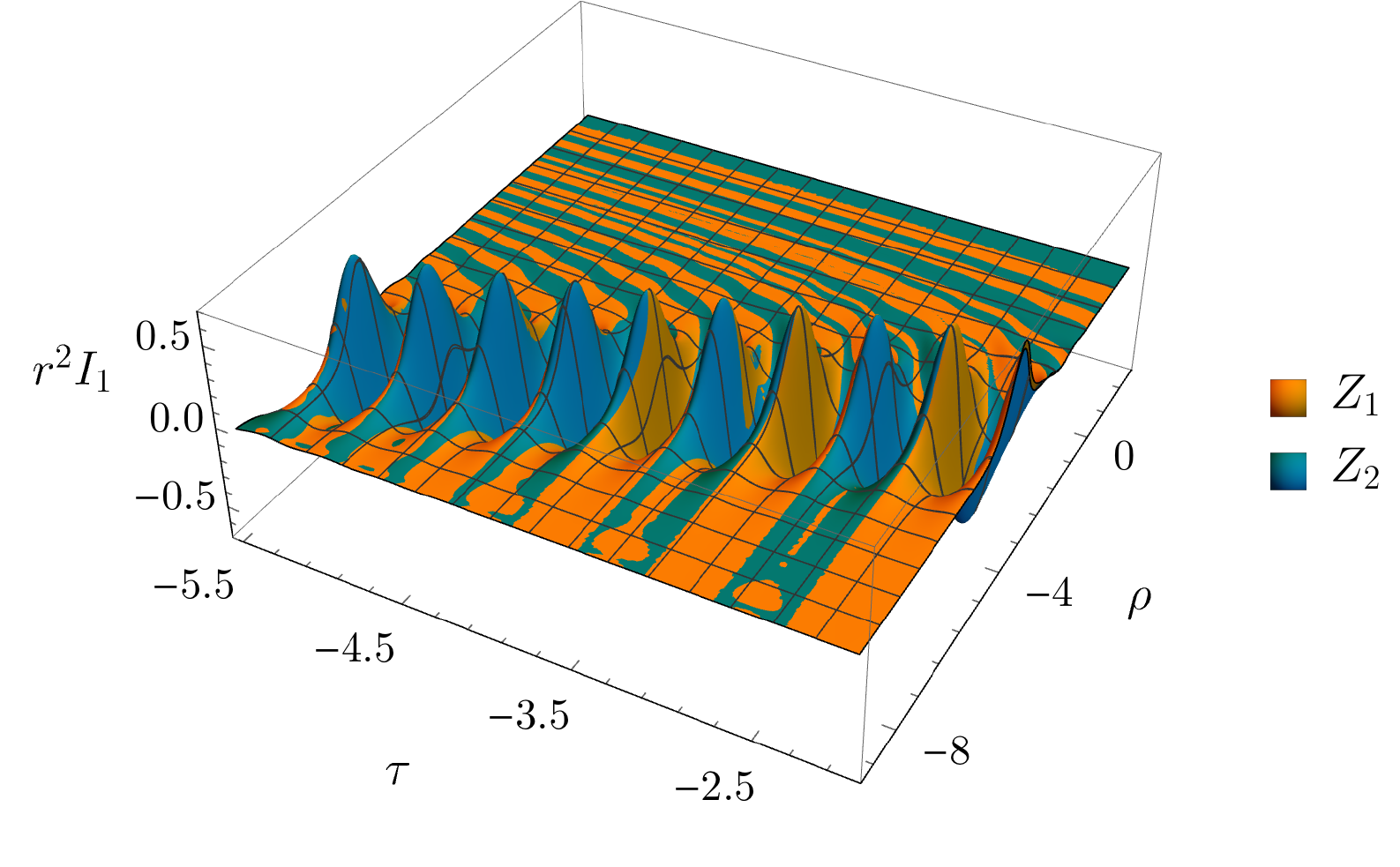}
    & \includegraphics[width=0.475\textwidth]{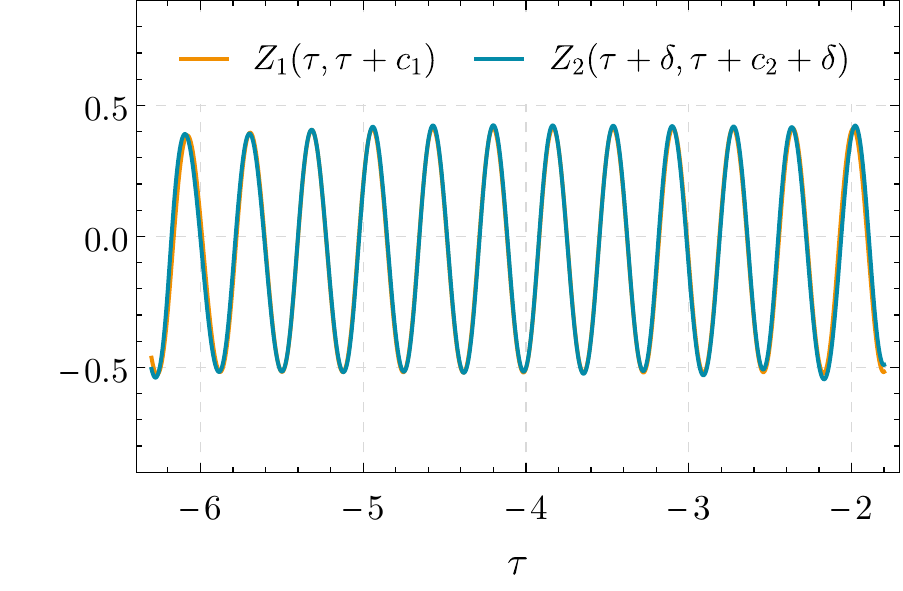}
  \end{tabular}
  \caption{Universality of the type II critical solution in the
    magnetic ansatz. We plot the spatio-temporal profile of the
    scale-invariant quantity $Z := r^2 I_1$, where the invariant $I_1$
    is defined in \eqref{eq:12} (note that the period is $\Delta/2$
    because this quantity is quadratic in dynamical variables). Two
    critical solutions were generated by a bisection search starting
    from the different initial conditions (i) and (iii) (for the
    latter we take $a_{1}=a_{2}=1$, $s_{1}=2$, $s_{2}=p$,
    $q_{1}=q_{2}=2$). Having two solutions $Z_1, Z_2$ expressed in
    terms of the coordinates $(\tau,\rho)$ defined in (\ref{eq:11}),
    we are allowed to perform any translation of one of them such that
    both coincide. (In practice we minimize the difference
    $\|Z_1(\tau,\rho)-Z_{2}(\tau+\delta_{1},\rho+\delta_{2})\|$ in
    some suitable norm over the shift parameters
    $(\delta_{1},\delta_{2})$.)  If the phenomenon is universal then
    both solutions should agree asymptotically as
    $(\tau,\rho)\rightarrow(-\infty,-\infty)$, which is demonstrated
    here.  The right plot shows the aligned profiles along the
    particular line $\tau-\rho=\const$ passing through the local
    extrema closest to the origin.}
  \label{f:MagneticCriticalSolution}
\end{figure*}

Universality of the critical solution is demonstrated in
Fig.~\ref{f:MagneticCriticalSolution}, where we compare
spatio-temporal profiles of solutions obtained through bisection
search starting from the different initial conditions (i) and (iii).
We do this by plotting a suitably rescaled invariant $I_{1}$ defined
in \eqref{eq:12} with respect to the coordinates (\ref{eq:11}).

\subsubsection{Type I collapse}

Next we turn to type I critical behavior, still in the magnetic
sector.  We consider initial data family (i) and vary $p := a_1$ in
\eqref{e:idw}, fixing the remaining parameters to $s_1 = 4$,
$x_1 = 10$ and $q_1 = 2$. The critical amplitude at the threshold
between dispersal and black hole formation is $p_{*}\approx -1.35232$.
As discovered in \cite{Choptuik1996}, the $n=1$ Bartnik-McKinnon
soliton $X_1$, which has one unstable mode in the magnetic sector,
plays the role of the critical solution. Our numerical simulations
reproduce this behavior.

\begin{figure}[!th]
  \centering
  \includegraphics[width=0.475\textwidth]{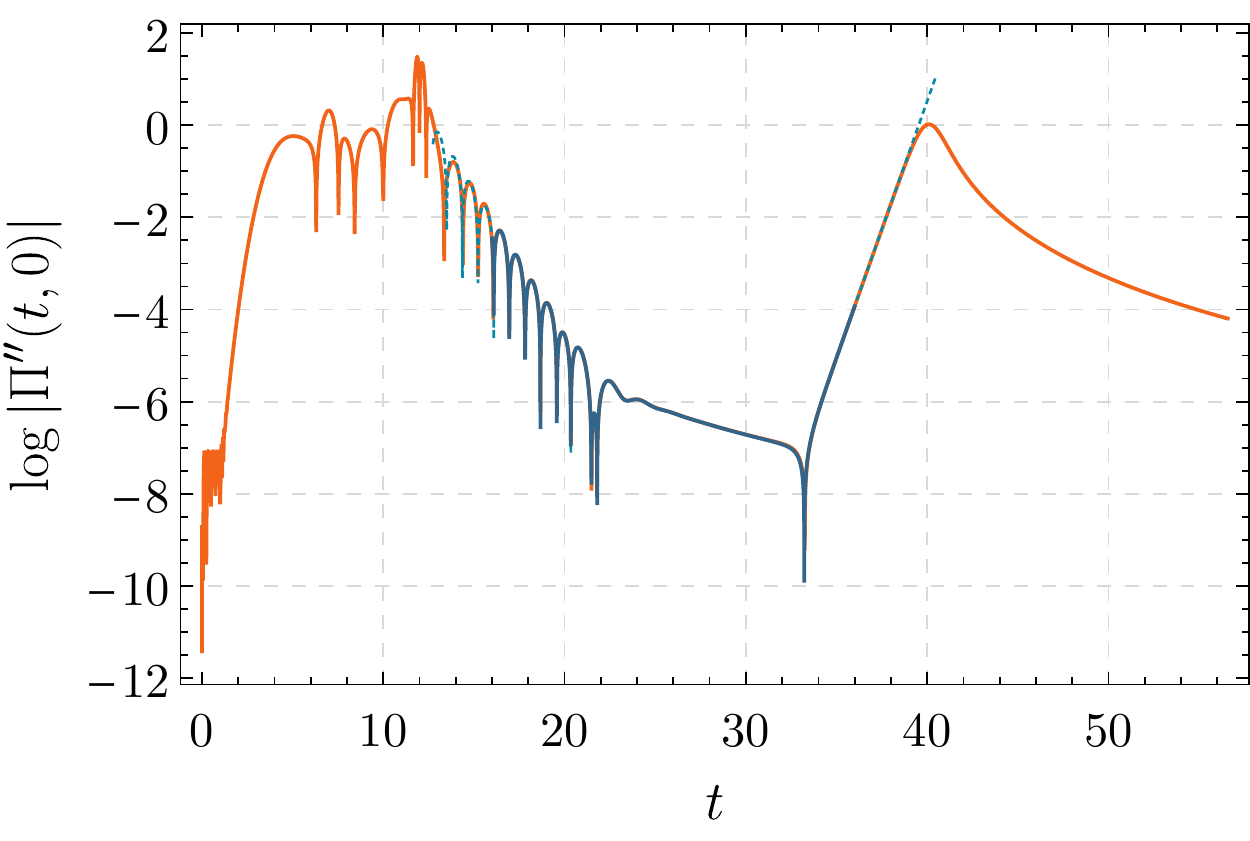}
  \caption{Subcritical evolution in type I critical collapse (orange)
    and the best-fit (blue; solid in the fitting range) of a linear
    combination of QNM, polynomial tail, and unstable mode of
    $X_{1}$. The fitting formula is
    $c_{1}\sin(\Omega (t-15)+c_{2})\exp(-\Gamma(t-15)) +
    c_{3}t^{p}\exp(c_{4}/t+c_{5}/t^{2})+c_{6}\exp(\lambda (t-34))$.
    The relevant parameters for this plot are $\Omega=3.639$,
    $\Gamma=1.426$, $p=-4.801$, and $\lambda=2.563$.}
  \label{f:MagneticConvergenceToX1}
\end{figure}

In addition, we investigate more closely how the dynamical solutions
approach the intermediate attractor. In Appendix \ref{s:linpert} we
carry out a linear perturbation analysis about $X_1$, which confirms
the unstable mode with exponent $\lambda \approx 2.56279$. In
addition, we have found quasinormal modes (QNM), the least damped of
which has $\lambda = -1.40233 \pm 3.60351 i$. Figure
\ref{f:MagneticConvergenceToX1} shows the different phases of the
evolution: approach to the unstable attractor $X_1$ via QNM and
polynomial tail, and departure along the unstable mode. The fitted
values of the QNM and unstable mode agree well with the prediction.
The tail does not appear for a sufficiently long time to allow for a
conclusive determination of the decay exponent $p$; our numerical fit
yields $p = -4.801$. (For comparison, the tail on a Schwarzschild or
Minkowski background has exponent $p=-4$ \cite{Bizon2007a}.)

As is characteristic of type I critical behavior, we observe a
saturation of the black hole mass in supercritical evolutions as a
function of the parameter distance from the critical solution. The
mass gap converges to the approximate value $0.5802$, which is close
to but slightly less than the mass of the $X_{1}$ solution
\cite{Bartnik1988}, $M_{X_{1}}=0.585942$. As the apparent horizon
forms, a fraction of the energy associated with $X_{1}$ stays outside
of the trapped region, and this excess of mass escapes to infinity
(however with our numerical code we are unable to follow this part of
the evolution).
\begin{figure}[!t]
  \centering
  \includegraphics[width=0.475\textwidth]{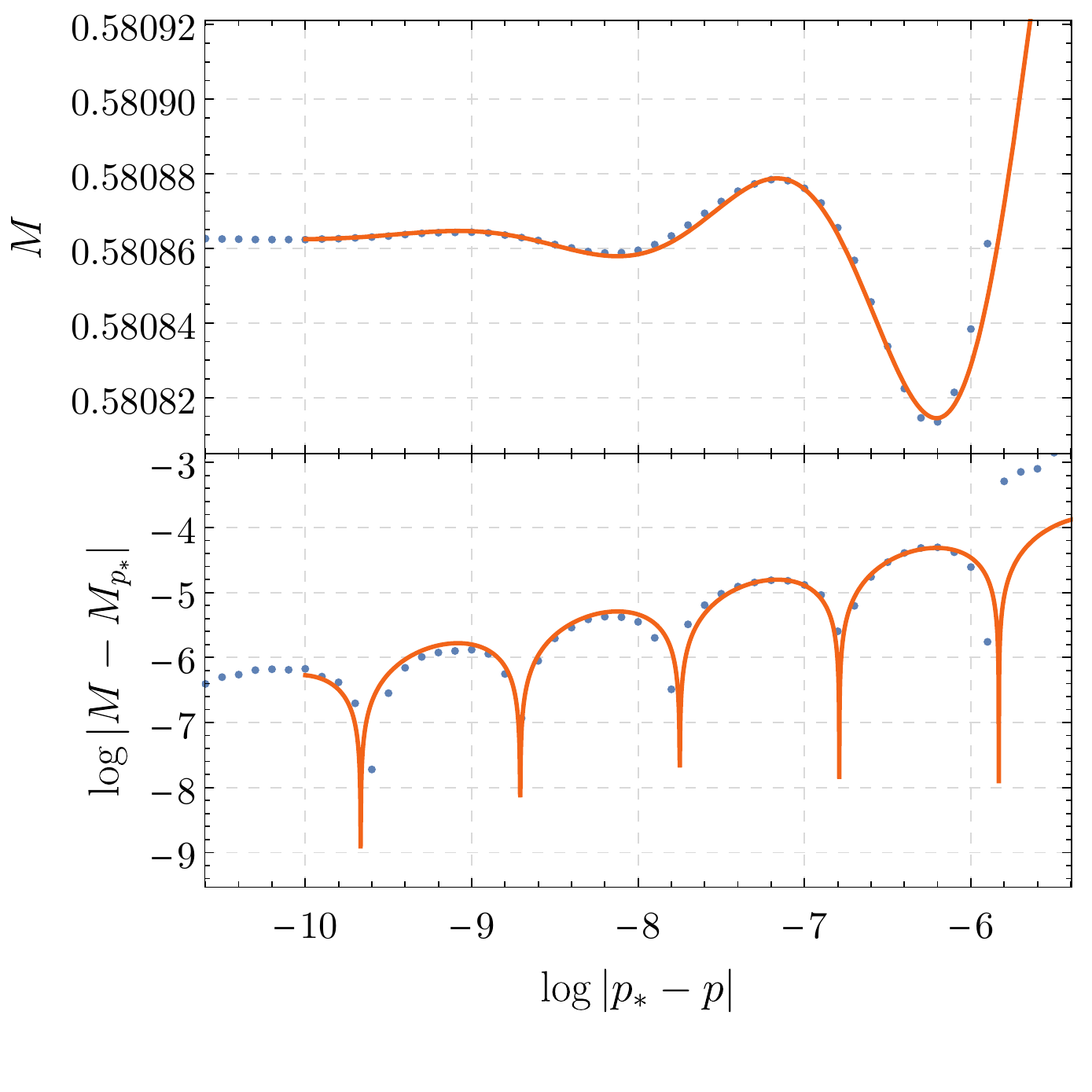}
  \caption{Mass of the apparent horizon in type I critical collapse as
    a function of the logarithm of the critical separation. For finite
    separation the mass oscillates around the asymptotic value
    $M_{p_{*}}$, which is slightly smaller than the mass of the
    critical solution $X_{1}$. This ``damped oscillation'' is
    suggested to be an imprint of the least damped QNM of $X_1$ (see
    the discussion in the text). The data points are plotted together
    with the fit
    $M_{p_{*}}+ c_{1}\cos\left(c_{2}\log|p_{*}-p| +c_{3}\right)
    \exp(c_{4} \log|p_{*}-p|)$, where $M_{p_{*}}$ and the $c_{i}$ are
    fitting parameters.}
  \label{f:MassOscillation}
\end{figure}

Moreover, on a plot of $M$ vs. $\log\left|p_{*}-p\right|$ we observe a
damped oscillation, see Fig.~\ref{f:MassOscillation}, whose origin may
be explained as follows.  As discussed above, the linear analysis of
$X_{1}$ predicts the existence of both stable and unstable modes. Thus
close to the critical point $p\approx p_{*}$ the dynamical solution
$w(t,r)$ consist of the attractor $w_{s}(r)$ and its linear
perturbation of the form
\begin{eqnarray}
  \label{eq:10}
  w(t,r) &=& w_{s}(r) + \phi_{UN}(r) \left|p_{*}-p\right|e^{\lambda t}\nonumber\\
  &&+ \phi_{QNM}(r) \sin\left(\Omega t\right)e^{-\Gamma t} + \cdots,
\end{eqnarray}
where $\lambda>0$ is the exponent of the unstable mode, the third term
represents the dominant QNM with $\Gamma>0$, and the dots represent
faster decaying modes and (possibly) the power law tail.  Performing
bisection in one parameter $p$, we effectively cancel the unstable
mode only but the magnitude of the QNM is not under control.
Therefore what contributes to the apparent horizon mass is the static
solution $X_{1}$ itself and its least damped QNM. Because the latter
oscillates (in time), its magnitude will depend on the time spent
close to $X_{1}$, which in turn depends on $\lambda$ and the distance
$|p_{*}-p|$ from the critical point. A simple calculation shows that
one should expect $M$ to oscillate (with respect to $|p_{*}-p|$) with
frequency $\Omega/\lambda$ and damping $\Gamma/\lambda$. However, from
the data we get numbers close to $2\Omega/\lambda$ and
$2\Gamma/\lambda$ for the frequency and damping respectively. This
suggests that the observed phenomenon is not a linear effect.


\subsection{Sphaleronic sector}

Next we switch on the sphaleronic sector in the general ansatz
\eqref{e:ymansatz} for the YM connection. For generic initial data
with $\omega \neq 0$ we observe type II critical behavior only. This
is not surprising because as explained in Sec.~\ref{s:intro}, the type
I critical solution in the magnetic sector, $X_1$, has an additional
unstable mode in the sphaleronic sector \cite{Lavrelashvili1995}.

\begin{figure}[t]
  \centering
  \includegraphics[width=0.437\textwidth]{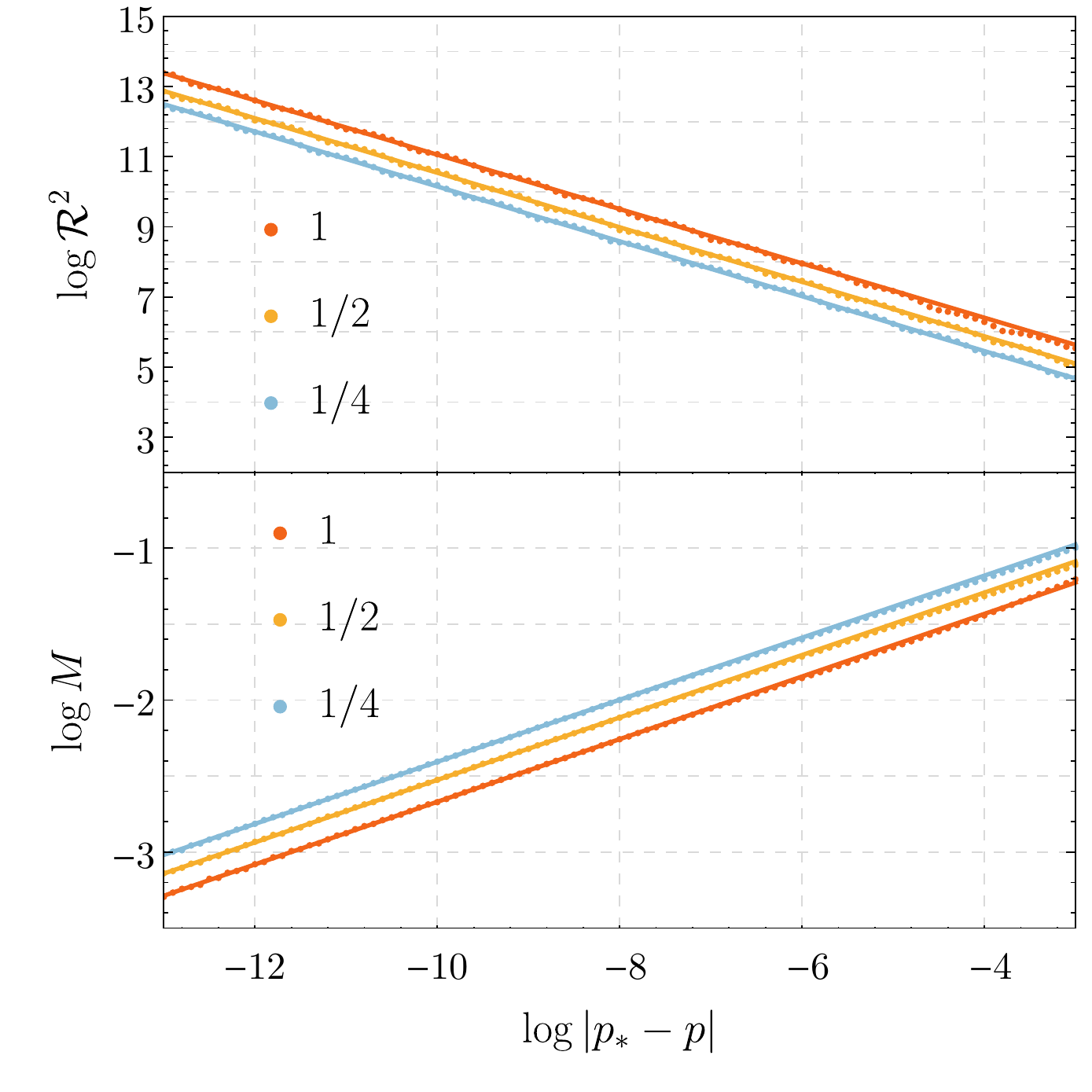}
  \caption{Supercritical (top) and subcritical (bottom) scaling
    characterizing type II critical collapse observed within the
    extended ansatz. The data are plotted with points together with
    best fits (lines). To produce the plot we used the family of
    initial data (ii) with bisection parameter $p:=s_2$, different
    values for $a_{1}=1, 1/2, 1/4$ (color-coded in the plot), and
    $a_{2}=\sqrt{a_{1}(2-a_{1})}$, $s_{1}=3$, $q_{1}=2$, $q_{2}=3$.
    In each case we find an exponent $\gamma$ close to the value in
    the magnetic sector both from super- and subcritical evolutions,
    see Table~\ref{tab:1}. The plot of $\log{\mathcal{R}^{2}}$
    vs. $\log\left|p_{*}-p\right|$ shows regular oscillations with
    period very close to the value found in the magnetic sector and
    consistent with the theoretical prediction
    (Sec.~\ref{s:typeI_II_mag}). Similar oscillations, though
    expected, are less noticeable and less regular on the lower plot
    due to insufficient resolution. (Precise determination of the
    location of the apparent horizon requires high resolution at a
    finite position.) Note the decimal logarithm is used on both
    plots.}
  \label{f:SphaleronicScalings}
\end{figure}

Figure~\ref{f:SphaleronicScalings} shows the sub- and supercritical
scaling of the black hole mass and Riemann curvature invariant for
different initial data from family (ii). The scaling exponents, shown
in Table~\ref{tab:1}, are close to the values in the magnetic sector
(see Sec.~\ref{sec:type-ii-collapse}) but deviate well beyond the
fitting error when the sphaleronic amplitude is increased.

\begingroup
\squeezetable
\begin{table}[h]
  \centering
  \begin{tabular}{c@{\qquad}c@{\qquad}c}
    \toprule
    $a_{1}$ & $\gamma$ supercritical & $\gamma$ subcritical \\ \colrule
    $1$ & $0.20612 \pm 0.00037$ & $0.19368 \pm 0.00088$ \\
    $1/2$ & $0.20545\pm 0.00031$ & $0.19431\pm 0.00098$ \\
    $1/4$ & $0.20422\pm 0.00028$ & $0.19558\pm 0.00092$ \\
    \botrule
    \end{tabular}
    \caption{Super- and subcritical scaling exponents $\gamma$, see
      (\ref{eq:15}-\ref{eq:9}), within the general ansatz for the
      family of initial data (ii) with parameters as for the data
      shown in Fig.~\ref{f:SphaleronicScalings}.  (The case $a_{1}=0$
      would correspond to the magnetic solution, since then also
      $a_2 = \sqrt{a_1(2-a_1)} = 0$.)}
  \label{tab:1}
\end{table}
\endgroup

\begin{figure}[!th]
  \centering
  \includegraphics[width=0.467\textwidth]{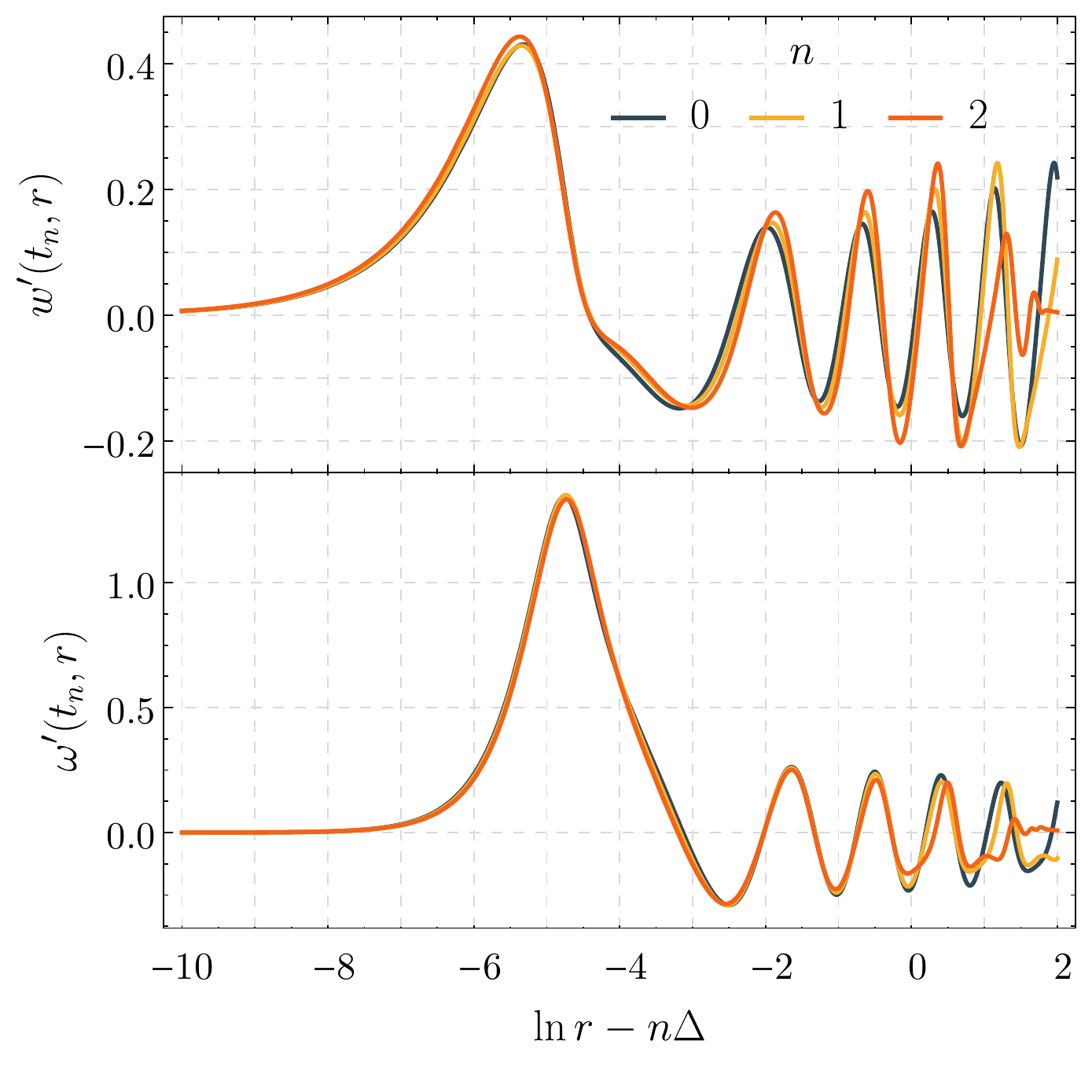}
  \hspace{3ex}
  \includegraphics[width=0.467\textwidth]{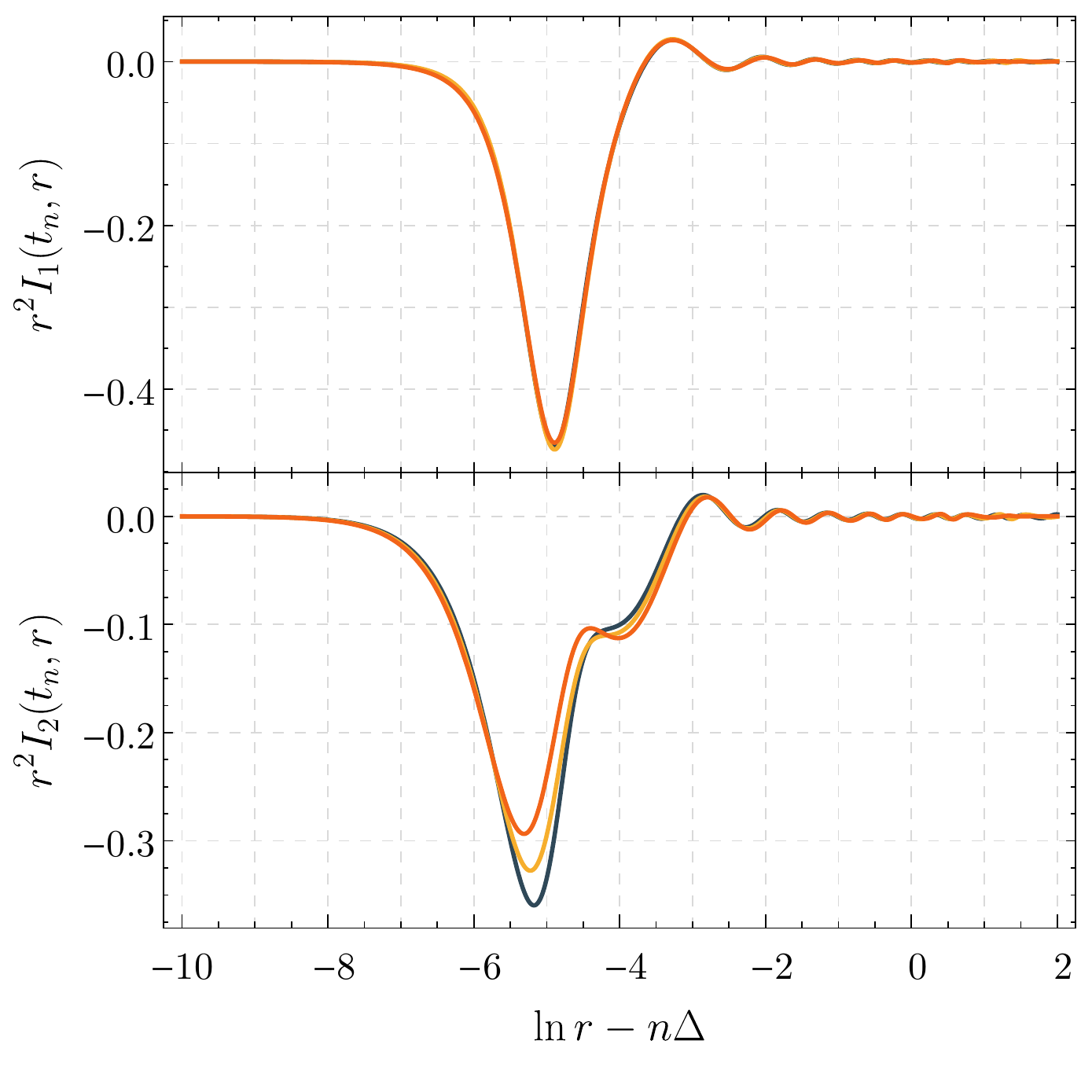}
  \caption{In the extended system neither $w'$ nor $\omega'$ appear to
    be exactly scale invariant (compare
    Fig.~\ref{f:MagneticPeriodicity}). However, the rescaled invariant
    $r^{2} I_{1}$ (\ref{eq:12}-\ref{eq:13}) does appear to be
    discretely self-similar; all presented functions are plotted at
    times selected to make the shifted profiles of $r^{2} I_{1}$
    overlap. We find $\Delta = 0.7445 \pm 0.0073$ for a solution
    constructed from the initial data (ii) with $a_{1}=a_{2}=1$,
    $s_{1}=2$, $s_{2}=p$, $q_{1}=2$, $q_{2}=3$. Observe however that
    the corresponding profiles of the rescaled second invariant
    $r^{2} I_{2}$ do \emph{not} overlap.  }
  \label{f:SphaleronicEchos_dw_domega}
\end{figure}

A close examination of the spatial profiles of the critical solution
shows that the quantities $w'$ and $\omega'$ are almost, but not
exactly, scale invariant (Fig.~\ref{f:SphaleronicEchos_dw_domega}).
In order to avoid potential gauge effects, we consider the manifestly
gauge invariant quantities $I_1$ (the Lagrangian) and $I_2$ defined in
\eqref{eq:13} and \eqref{eq:14}. Figure
\ref{f:SphaleronicEchos_dw_domega} indicates that while the profiles
of $I_1$ can be made to overlap, those of $I_2$ at the corresponding
times do not. Thus our solution is not exactly self-similar. In any
case, from $I_1$ we extract an echoing exponent
$\Delta = 0.7445 \pm 0.0073$ consistent with the value in the magnetic
sector.

\begin{figure*}[!ht]
  \centering
  \begin{tabular}{c}
    \includegraphics[width=0.475\textwidth]{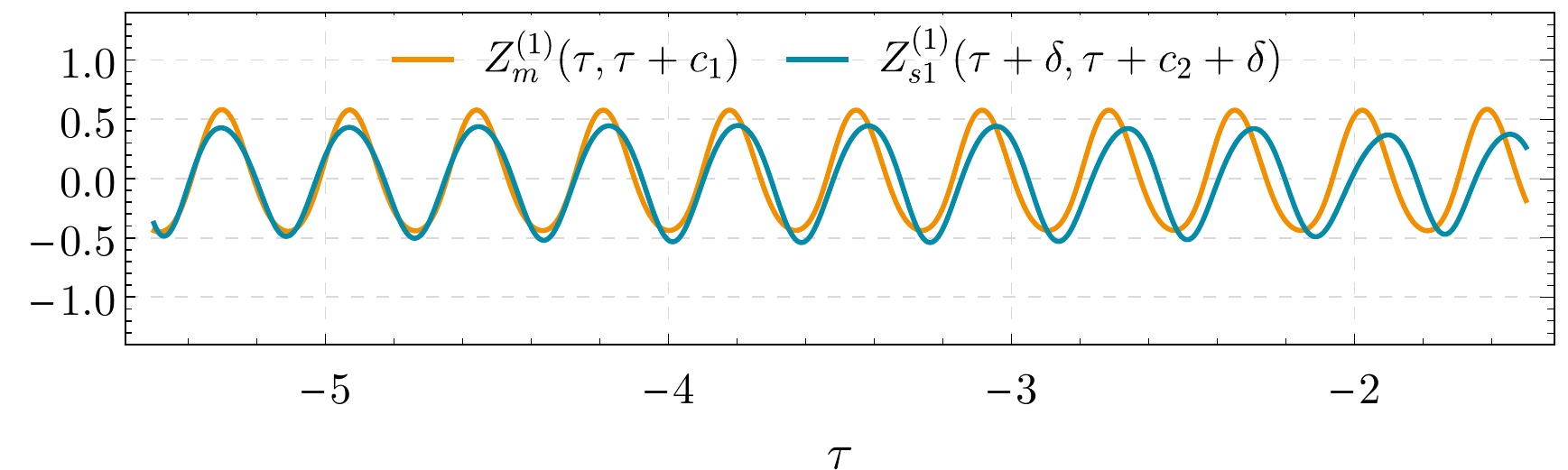}
    \\
    \includegraphics[width=0.475\textwidth]{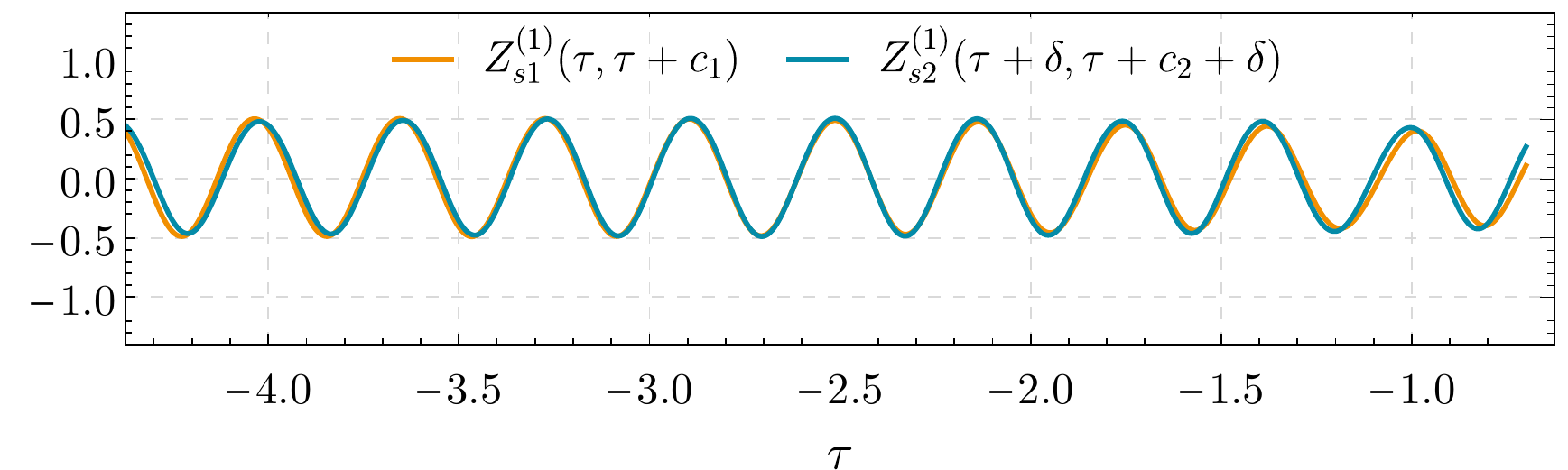}
    \\
    \includegraphics[width=0.475\textwidth]{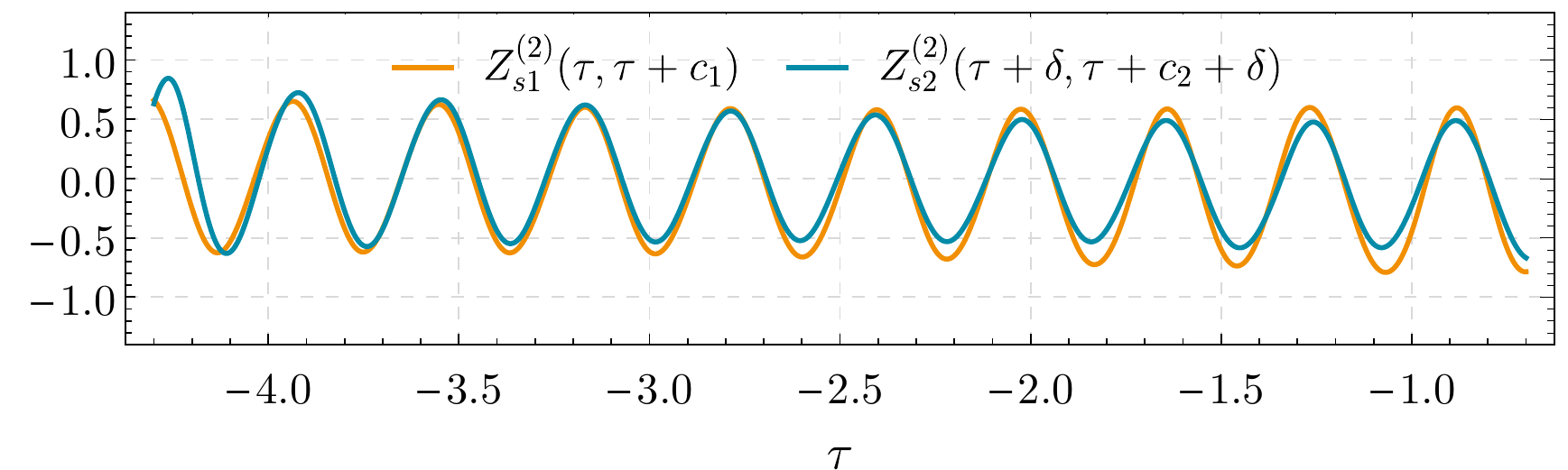}
  \end{tabular}
  \caption{Comparison of type II critical solutions. We plot the
    rescaled invariants $Z^{(1)} := r^2 I_1$ and
    $Z^{(1)} := r^{2}I_{2}$. In the upper plot we present the first
    invariant, where $Z_{m}$ refers to the magnetic critical solution
    obtained in Sec.~\ref{s:typeI_II_mag}, i.e. we take initial data
    (i) with $p:=a_{1}$ and $s_{1}=1/4$, $x_{1}=3$, $q_{1}=1$, whereas
    $Z_{s1}$ was obtained from sphaleronic initial data (ii) with
    $p:=s_{2}$ and $a_{1}=a_{2}=1$, $s_{1}=1$, $q_{1}=2$,
    $q_{2}=3$. Note that $Z^{(2)}_{m}\equiv 0$. The middle and bottom
    plots show two sphaleronic solutions with different initial
    conditions: family (ii) with $p=s_2$, $a_{1}=a_{2}=1$, $q_{1}=2$,
    $q_{2}=3$ for both solutions, but $s_{1}=2$ for $Z_{s1}$ and
    $s_{1}=1$ for $Z_{s2}$. As in
    Fig.~\ref{f:MagneticCriticalSolution} we show the aligned profiles
    along the line $\tau-\rho=\const$ passing through the local
    extrema closest to the origin.}
  \label{f:SphaleronicCriticalSolution}
\end{figure*}

In Fig.~\ref{f:SphaleronicCriticalSolution} we compare the invariants
of the magnetic critical solution with those of the sphaleronic one.
There is no exact agreement for the first invariant $I_1$.  Moreover,
while the second invariant $I_2$ is identically zero in the magnetic
sector, it is comparable in amplitude to $I_1$ in the sphaleronic
sector. This indicates that the two critical solutions might not be
the same. Figure \ref{f:SphaleronicCriticalSolution} also indicates
that there is no perfect universality: $I_1$ for the critical
solutions from two different initial data in the extended system shows
reasonably good agreement but $I_2$ does not.

Our preliminary conclusion is that there are indications that the type
II critical solutions in the magnetic sector and in the sphaleronic
sector might not be identical, that the sphaleronic critical solution
might not be exactly discretely self-similar, and that exact
universality might be lost. We did investigate whether these findings
might be the caused by numerical errors but could not see any signs of
significantly worse convergence of the numerical solution in the
sphaleronic sector as compared with the magnetic sector.

\begin{figure}[!th]
  \centering
  \includegraphics[width=0.467\textwidth]{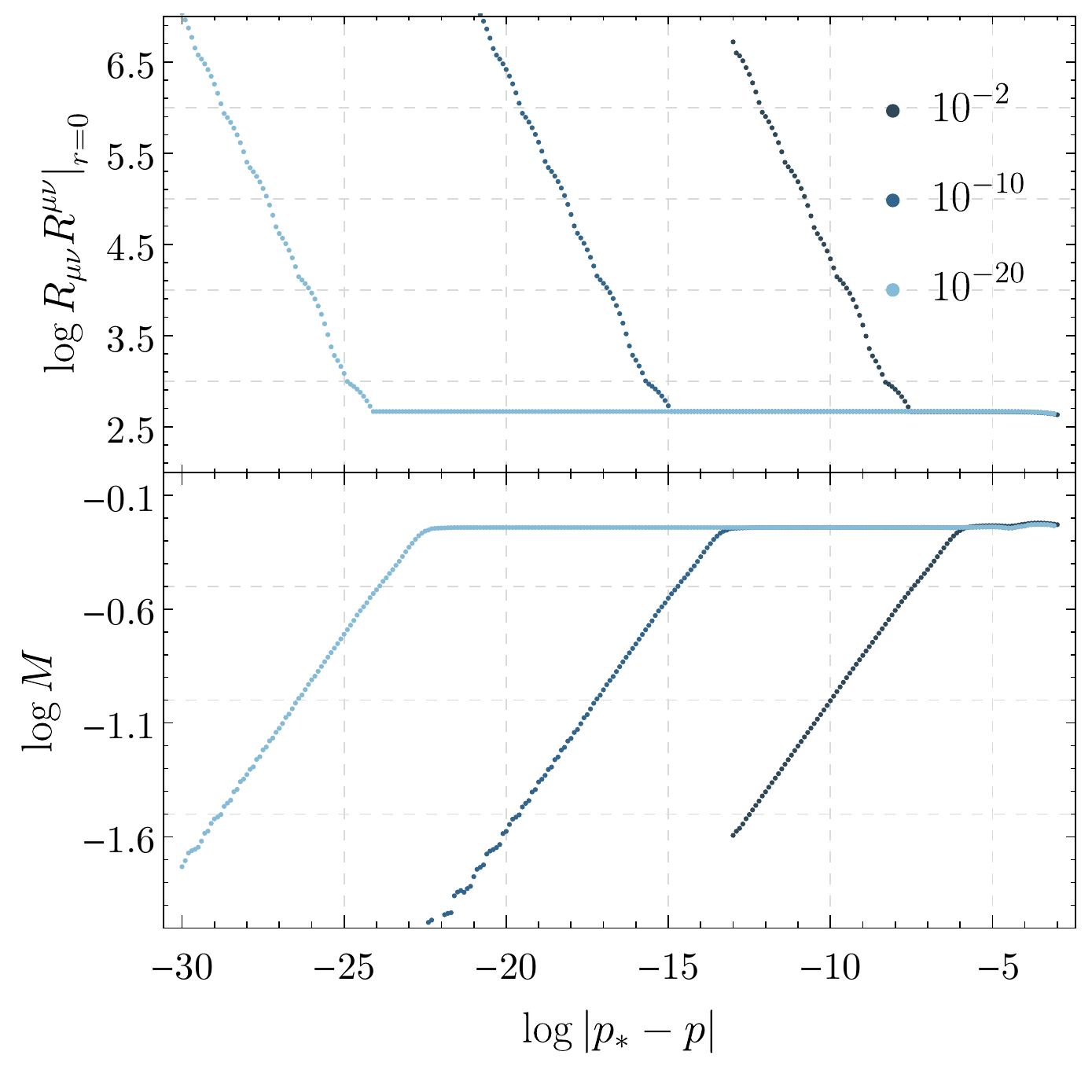}
  \caption{For initial data (i) the type II critical behavior sets off
    at a finite distance from the critical point. The onset of the
    polynomial scaling depends on the strength of the sphaleronic
    perturbation. The legend shows the amplitude $a_2$ of $\omega$ in
    initial data class (i). The bisection parameter is $p:=a_1$ and
    the other parameters are fixed to $s_{2}=1/4$, $x_{2}=1$ and
    $q_{2}=1$.}
  \label{f:SphaleronicScalingsMulti}
\end{figure}

We shall leave this question aside for the time being and look more
closely at how the type I critical behavior seen in the magnetic
sector is transformed into type II behavior when the sphaleronic
perturbation is turned on. This is demonstrated in
Fig.~\ref{f:SphaleronicScalingsMulti}, where we consider the family of
initial data (i). We take the same set of parameters as used to
produce Fig.~\ref{f:MagneticConvergenceToX1}, but in addition we
include a small sphaleronic amplitude $a_2$, while the bisection
parameter is still $p := a_1$. The smaller the sphaleronic
perturbation, the closer one needs to tune to the critical point in
order to see the polynomial scaling of the mass and curvature
invariant characteristic of type II behavior.

\begin{figure}[!th]
  \centering
  \includegraphics[width=0.475\textwidth]{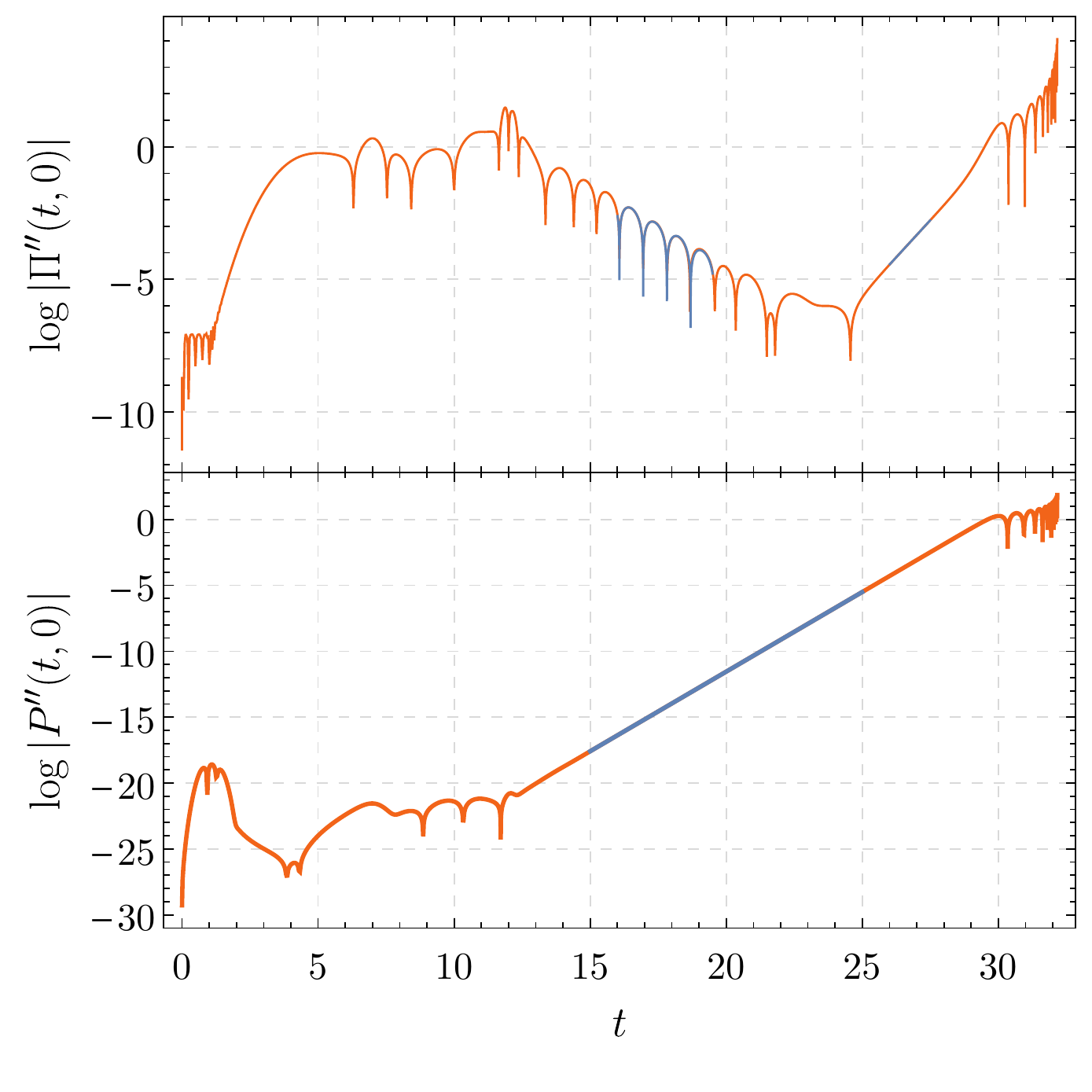}
  \caption{The Bartnik-McKinnon soliton $X_{1}$ as an intermediate
    attractor of a near-critical evolution in the extended
    ansatz. Here the sphaleronic perturbation $\omega$ in the initial
    data was held fixed at $a_{2}=10^{-20}$ and the amplitude $p:=a_1$
    of the initial data for $w$ was tuned to criticality (initial data
    class (ii), the same as used for
    Fig.~\ref{f:SphaleronicScalingsMulti}). Since this procedure
    controls only one of the two unstable modes of $X_1$, this static
    solution only appears as an intermediate attractor. Ultimately the
    evolution drifts away from $X_{1}$ and echoes of the discretely
    self-similar type II solution become visible. Compare with
    Fig.~\ref{f:MagneticConvergenceToX1}.}
  \label{f:SphaleronicConvergenceToX1}
\end{figure}

Even though $X_1$ is not a critical solution in the extended ansatz,
it nevertheless plays the role of an intermediate attractor for data
close to type I critical data in the magnetic ansatz with a small
sphaleronic perturbation. This is illustrated in
Fig.~\ref{f:SphaleronicConvergenceToX1}, where $w$ decays to $X_1$ by
the dominant QNM before it departs along the unstable mode, whereas
$\omega$ only shows an unstable mode.  The fitted exponents agree well
with the predictions from linear perturbation theory (Appendix
\ref{s:linpert}), given in brackets:
$\lambda \approx -1.41995 \pm 3.60267 i \, (-1.40233\pm 3.60351i)$ for
the QNM, $\lambda \approx 2.57355 \, (2.56280)$ for the unstable mode
in $w$ and $\lambda\approx 2.78296 \, (2.78310)$ for the unstable mode
in $\omega$.


\section{Type III collapse}
\label{s:typeIII}

In this section we present our numerical results on type III critical
behavior. These simulations were carried out using the code based on
CMC-isotropic coordinates; the value of the mean extrinsic curvature
of the slices is taken to be $K=1/2$.


\subsection{Initial data}

We use the same family of initial data as in \cite{Rinne2014a}
consisting of a ``kink'' and a ``bump'' in $w$, and a ``bump'' in
$\omega$:
\begin{align}
  \label{e:idw_type3}
  w(0,r) &= - \tanh \left(\frac{r - r_k}{\sigma_k}\right) - A_b \exp \left(
  - \frac{(r - r_b)^2}{2 \sigma_b^2} \right),\\
  \label{e:idomega_type3}
  \omega(0,r) &= \tilde A_b \exp \left(
  - \frac{(r - \tilde r_b)^2}{2 \tilde \sigma_b^2} \right),\\
  \dot w(0,r) &= \dot \omega(0,r) = 0.
\end{align}


\subsection{Magnetic sector}

We begin by restricting ourselves to the magnetic sector, i.e. we set
$\tilde A_b = 0$ in \eqref{e:idomega_type3}.  We fix $r_k = 0.4$,
$r_b = 0.7$ and $\sigma_b = \sigma_k = 0.05$ in \eqref{e:idw_type3}
and vary $A_b$.  The critical amplitude is found to be
$A_b^* = 1.2539174811047301$.

\begin{figure}
\centerline{\includegraphics[width=.475\textwidth]{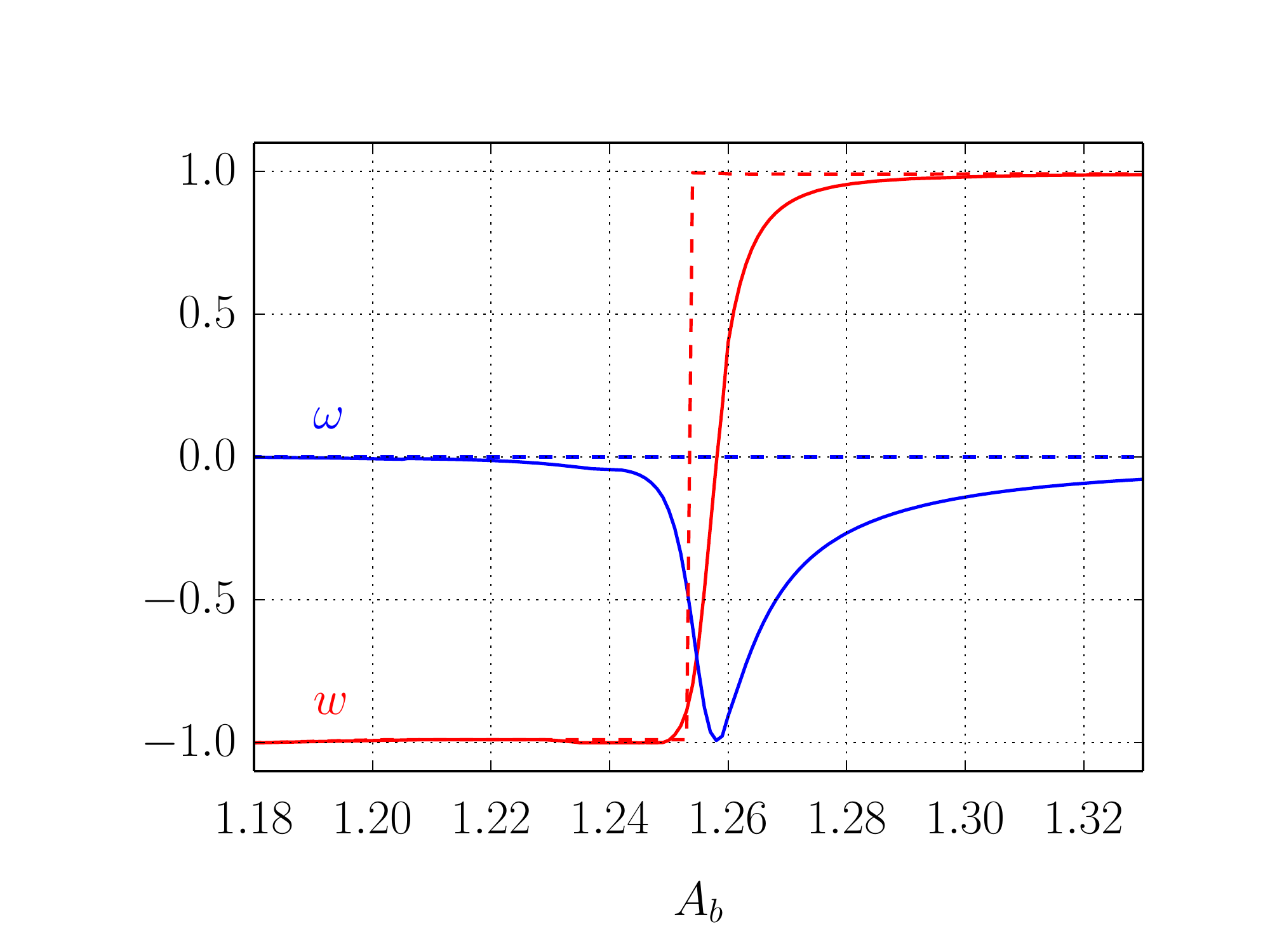}}
\caption{\label{f:type3_w_wtilde}
  Final values of $w$ (red) and $\omega$ (blue) as functions of $A_b$
  in the magnetic sector ($\tilde A_b = 0$, dashed lines)
  and with a sphaleronic perturbation ($\tilde A_b = 10^{-2}$, solid lines).
}
\end{figure}

\begin{figure}
\centerline{\includegraphics[width=.475\textwidth]{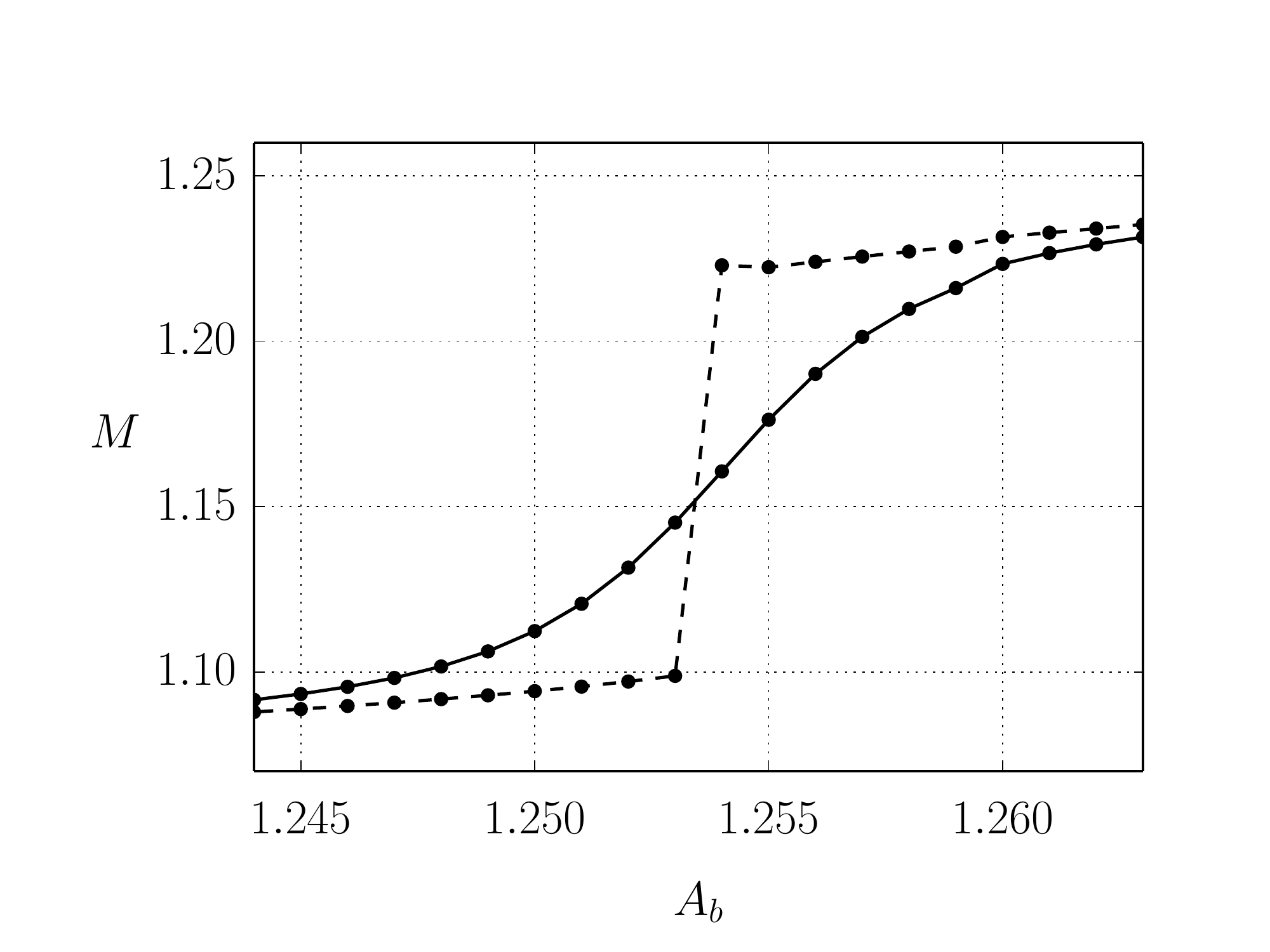}}
\caption{\label{f:type3_M}
  Mass $M$ of the final Schwarzschild black hole as a function of $A_b$
  in the magnetic sector ($\tilde A_b = 0$, dashed line)
  and with a sphaleronic perturbation ($\tilde A_b = 10^{-2}$, solid line).
}
\end{figure}

The results of our critical search confirm what was dubbed type III
critical collapse in \cite{Choptuik1999}. The final states of the
evolutions are Schwarzschild black holes with either $w=1$ or $w=-1$
(dashed line in Fig. \ref{f:type3_w_wtilde}), both of which correspond
to vacuum.  At the threshold between the two outcomes, the $n=1$
colored black hole \cite{Bizon1990,Volkov1990}, $Y_1$, is approached
as a codimension-one unstable attractor. This solution has one
continuous parameter, the horizon (areal) radius, which has the value
$2.11$ in our case.  The masses of the final Schwarzschild black holes
as the threshold is approached from either side are different
(Fig. \ref{f:type3_M}): in our case $M=1.235$ for $w=1$ and $M=1.090$
for $w=-1$.  The dependence of the mass gap on the horizon radius of
$Y_1$ was studied in detail in \cite{Rinne2014a}.

\begin{figure}
\centerline{\includegraphics[width=.475\textwidth]{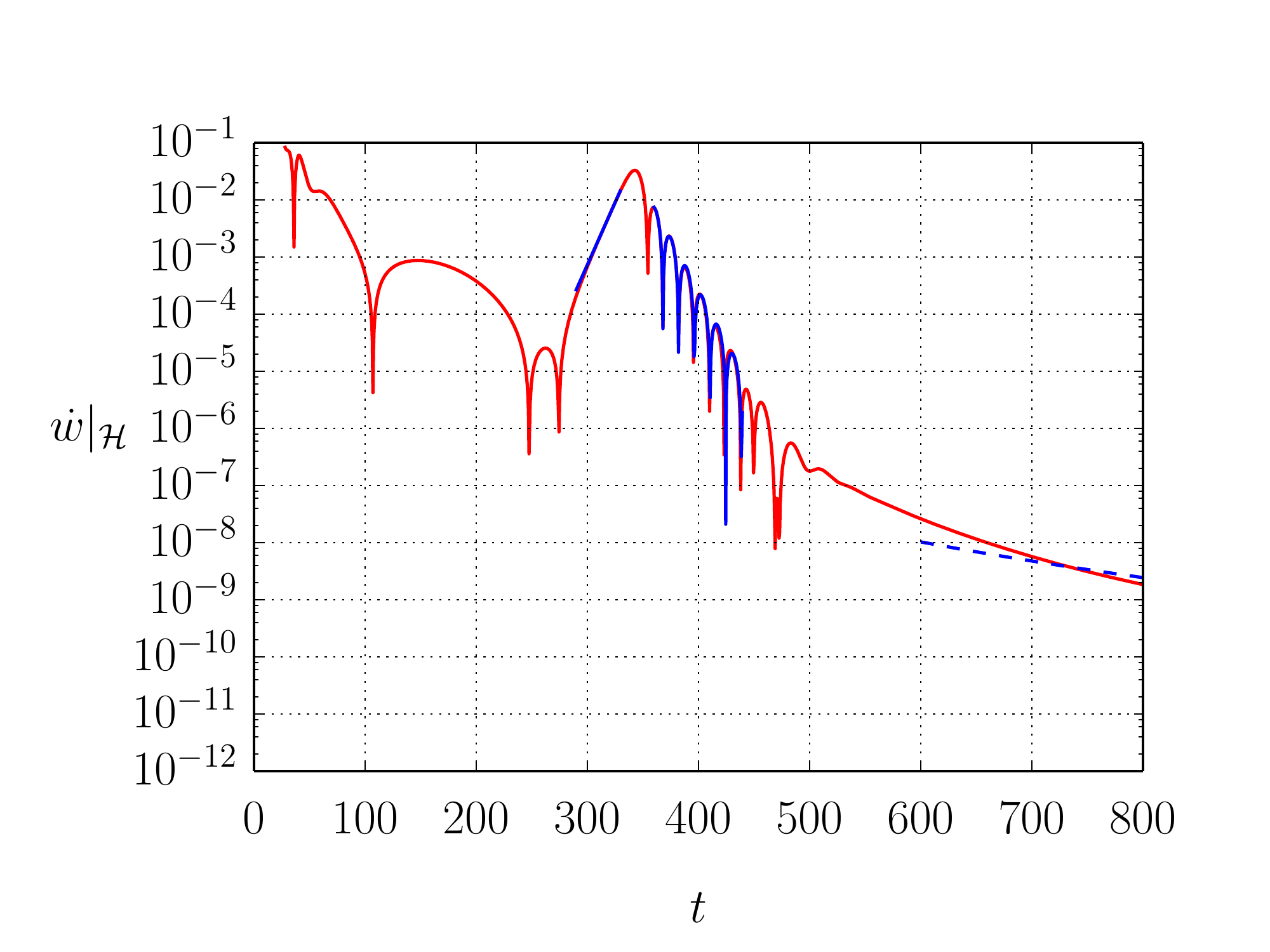}}
\caption{\label{f:type3_mag_dtw}
  Time derivative of $w$ at the horizon (after it forms) as a function
  of time for $\tilde A_b = 0$ (magnetic sector) and $A_b$ tuned to
  criticality (with final value $w=-1$ in this evolution).
  The solid blue curves are the fits to the unstable mode of $Y_1$
  and the QNM of the final Schwarzschild black hole.
  The dashed blue curve indicates the expected decay exponent ($p=-4$)
  of the tail, which is not attained here due to a lower resolution
  used in the critical bisection search.
}
\end{figure}

Figure~\ref{f:type3_mag_dtw} shows the different phases of a
near-critical evolution: decay $Y_1$ via QNM, departure along the
unstable mode of $Y_1$, and ringdown to the final Schwarzschild
solution via QNM and tail. Unlike for $X_1$
(Sec.~\ref{s:typeI_II_mag}), for $Y_1$ the period of oscillation of
the QNM is large compared to the timescale of the unstable mode so
that we only see one or two oscillations; this makes a fit difficult.
The fitted value of the unstable mode of $Y_1$, $\lambda = 0.1007$,
agrees well with the value $\lambda = 0.1020$ computed from linear
perturbation theory in \cite{Rinne2014a}. For the final ringdown to
Schwarzschild spacetime, the fitted value of the QNM frequency
$\lambda = -0.0835 \pm 0.2222 i$ matches the prediction
$\lambda = -0.0848 \pm 0.2278 i$ from linear perturbation theory
(\cite{Bizon2010a}, note the QNM frequency scales with $M^{-1}$, here
$M=1.090$).  The tail could not be resolved properly here due to a
lower resolution used in the time-consuming critical bisection search;
however for a higher resolution using the same code, the expected
\cite{Bizon2007a} exponent $p=-4$ was found in \cite{Rinne2014a}, and
we will observe the same exponent below in the sphaleronic evolutions.


\subsection{Sphaleronic sector}

Next, we add a small perturbation in $\omega$ to the initial data: we
choose $\tilde A_b = 10^{-2}$, $\tilde r_b = 0.7$ and
$\tilde \sigma_b = 0.05$. The discontinuous transition in $w$ as we
vary $A_b$ is now replaced by a continuous one, and the final $\omega$
also varies continuously (solid lines in Fig. \ref{f:type3_w_wtilde}).
The mass gap also disappears (solid line in Fig. \ref{f:type3_M}).

These findings are not surprising because the dichotomy between the
vacua $w=\pm 1$ in the magnetic sector is replaced by a continuum of
vacua
\begin{equation}
  w^2 + \omega^2 = 1
\end{equation}
in the general system. Hence it is impossible to perform a critical
search between two different outcomes. Moreover, as pointed out in
Sec.~\ref{s:intro}, $Y_1$ has an additional unstable mode in the
sphaleronic sector \cite{Volkov1995}, hence it cannot appear as a
critical solution in the extended system.

\begin{figure}
\centerline{\includegraphics[width=.475\textwidth]{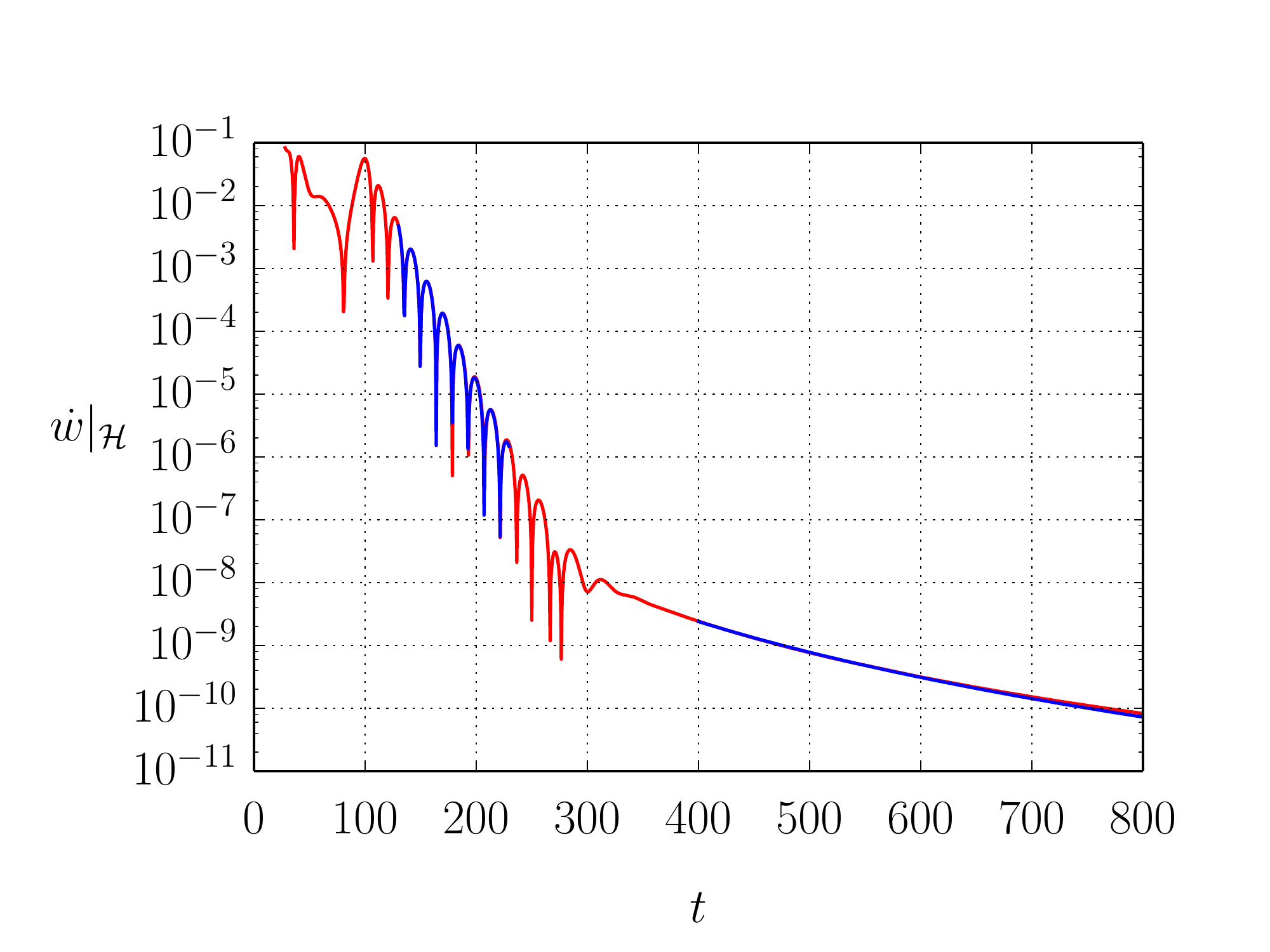}}
\caption{\label{f:type3_sph_dtw}
  Time derivative of $w$ at the horizon (after it forms)
  as a function of time for $\tilde A_b = 10^{-2}$ and the same value
  for $A_b$ as in the magnetic sector evolution (Fig. \ref{f:type3_mag_dtw}).
  The blue curves indicate the fits to the QNM and tail.
}
\end{figure}

\begin{figure}
\centerline{\includegraphics[width=.475\textwidth]{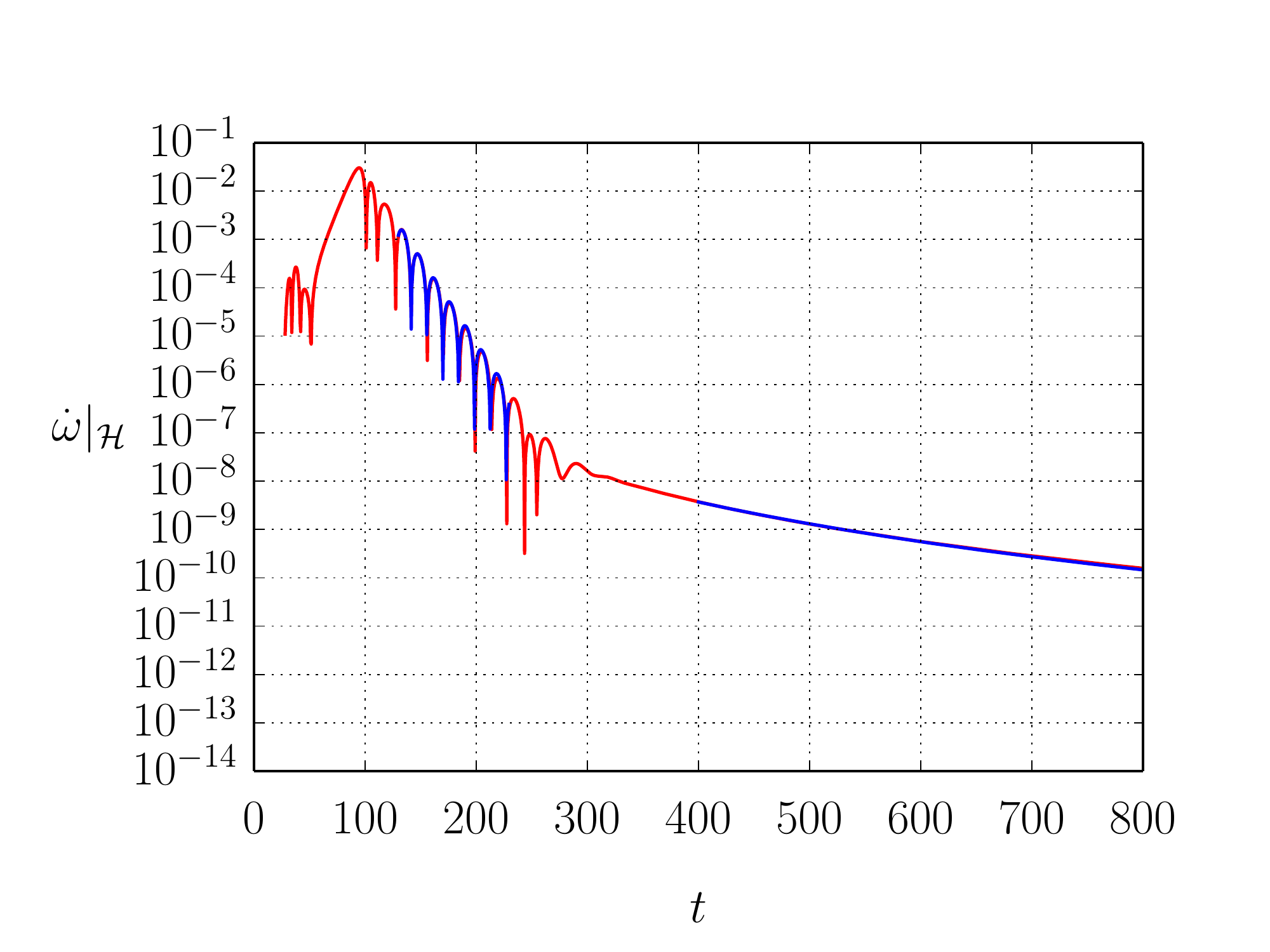}}
\caption{\label{f:type3_sph_dtomega}
  Time derivative of $\omega$ at the horizon (after it forms)
  as a function of time for $\tilde A_b = 10^{-2}$ and the same value
  for $A_b$ as in the magnetic sector evolution (Fig. \ref{f:type3_mag_dtw}).
  The blue curves indicate the fits to the QNM and tail.
}
\end{figure}

Figures \ref{f:type3_sph_dtw} and \ref{f:type3_sph_dtomega} show the
dynamical evolution for the value of $A_b$ that corresponded to the
critical threshold in the magnetic sector, but now with the
sphaleronic perturbation added. The QNM ringdown to an intermediate
attractor is no longer visible, only the QNM and tail to the final
Schwarzschild black hole. A fit to the QNM yields
$\lambda = -0.0819 \pm 0.2188 i$ for $w$ and
$\lambda = -0.0801 \pm 0.2204 i$ for $\omega$. This is to be compared
with the predicted QNM frequency in the magnetic sector
\cite{Bizon2010a} for the same mass (here $M=1.139$),
$\lambda = -0.0812 \pm 0.2180 i$. Thus our results support the claim
that the dominant Schwarzschild QNM frequency in the full EYM system
is the same as in the magnetic sector. The fitted tail exponents are
$p=-4.05$ for $w$ and $p=-3.66$ for $\omega$, both consistent with the
exponent $p=-4$ observed in the magnetic sector.


\section{Conclusions}
\label{s:concl}

This paper studies critical collapse in the general spherically
symmetric Einstein-Yang-Mills (EYM) system. Compared to the magnetic
ansatz most often used in numerical work so far, this has an
additional physical degree of freedom, the ``sphaleronic sector.''
Our main results can be summarized as follows.

In the magnetic sector, we confirm the phenomenology reported in
\cite{Choptuik1996}: both type I and type II critical collapse appear,
depending on the family of initial data chosen. In addition to
previous results, we find periodic wiggles in the type II scaling of
the Ricci curvature invariant in subcritical evolutions that we relate
to the echoing exponent. In type I collapse, our dynamical numerical
evolutions show an approach to the static critical solution, the
Bartnik-McKinnon soliton $X_1$, via a quasinormal mode (QNM) and a
tail. We compare this with a calculation of the QNM of $X_1$ in linear
perturbation theory. This is one of the few examples where a QNM
ringdown to a nontrivial unstable static solution has been studied
(other examples being the YM evolutions on a fixed Schwarzschild
background in \cite{Bizon2010a} and on the extremal
Reissner-Nordstr{\"{o}}m black hole in \cite{Bizon2016}). The presence
of the QNM also causes damped oscillations of the apparent horizon
mass as a function of the critical parameter distance in type I
collapse (Fig.~\ref{f:MassOscillation}).

When the sphaleronic sector is turned on in the initial data, the
picture of critical collapse changes completely. The type I behavior
now disappears and the generic critical behavior is type II. This is
not surprising as the magnetic critical solution $X_1$ has an
additional unstable mode in the sphaleronic sector
\cite{Lavrelashvili1995}. The supercritical mass and subcritical
curvature scaling exponents are very close to but, depending on the
initial data, not identical with the ones found in the magnetic
sector. We present a detailed comparison of the critical solution in
the extended system with the critical solution in the magnetic ansatz.
Looking at gauge invariant quantities $I_{1}$ and $I_{2}$ (see
(\ref{eq:12}) and (\ref{eq:13}) for their definition) indicates that
the two critical solutions are probably not identical. This follows
from the observation that $I_{2}$ is nonzero (comparable in size to
$I_{1}$) for critical evolutions of type II in the general ansatz,
whereas it vanishes identically in the purely magnetic sector. We also
find tentative evidence that exact discrete self-similarity as well as
universality of the critical solution (with regard to different
families of initial data) might be lost in the extended system.  It
could be that we are not yet sufficiently close to the critical point
to see the true features of the critical solution. However, to push
the bisection search further, we would have to use higher than the
native double precision and in addition increase the numerical
resolution much further, which did not seem feasible currently.

When a sphaleronic perturbation is added to initial data that would be
type I critical in the magnetic sector, the type II polynomial scaling
sets off at a finite distance from the critical point depending on the
strength of the sphaleronic perturbation
(Fig.~\ref{f:SphaleronicScalingsMulti}). In such evolutions the
magnetic type I critical solution $X_1$ can be seen as an intermediate
attractor before the type II attractor is approached. We observe a QNM
ringdown to this intermediate attractor, and again we find good
agreement of the QNM frequency with a calculation in linear
perturbation theory.

There is a third type of critical collapse in the magnetic sector of
the EYM system discovered in \cite{Choptuik1999} (and recently studied
in more detail in \cite{Rinne2014a}). Here evolutions on both sides of
the threshold eventually settle down to Schwarzschild black holes but
the YM potential is in different vacuum states. The critical solution
is the colored black hole $Y_1$. Our simulations give tentative
evidence of a QNM ringdown to the critical solution $Y_1$ but the time
range during which this becomes visible is too short to be able to fit
the QNM frequency. Higher precision would be required to uncover the
QNM ringdown as well as possibly a polynomial tail around this
intermediate unstable attractor. An independent confirmation of the
existence of QNMs of colored black holes and their spectra will
require a detailed analysis (boundary conditions) of the linearized
problem.

When a sphaleronic perturbation is included in the initial data, the
discontinuous transition of the YM potential $w$ and the final black
hole mass across the critical threshold is replaced with continuous
ones. Thus we can no longer tune the initial data between two distinct
outcomes, and the type III critical phenomenon disappears. This can be
explained by the existence of an additional unstable mode of the $Y_1$
critical solution in the sphaleronic sector \cite{Volkov1995}.


\begin{acknowledgments}
  The authors would like to thank Peter Aichelburg, Piotr Bizo\'n and
  Piotr Chru\'sciel for many helpful discussions. We also thank
  Carsten Gundlach for valuable remarks and comments. M.M. gratefully
  acknowledges the support of the Alexander von Humboldt Foundation
  and the Polish National Science Centre grant
  No.~DEC-2012/06/A/ST2/00397. The early stages of O.R.'s work on this
  project were supported by a Heisenberg Fellowship and Grant No.~RI
  2246/2 of the German Research Foundation (DFG). Computations were
  performed on the supercomputer Datura operated by AEI Potsdam.
\end{acknowledgments}


\appendix

\section{Field equations}
\label{s:fieldeqns}

In this appendix we present the formulations of the field equations
used in our two independent codes. The code used to study type I and
type II collapse combines polar-areal coordinates
\eqref{e:ds2_polar-areal} with the parametrization \eqref{e:ymansatz}
of the YM connection. The code used to study type III collapse employs
CMC-isotropic coordinates \eqref{e:ds2_CMC} and the parametrization
\eqref{e:ymansatz2} of the YM connection. The choice of different
parametrizations is insignificant and is only for ``historical
reasons'' in the development of our codes. The EYM equations for a
general spherically symmetric metric were also derived in the appendix
of \cite{Choptuik1999} and are consistent with our formulations.  We
use units in which $4\pi G g^{-2} = 1$, where $G$ is Newton's constant
and $g$ is the YM coupling constant. Throughout an overdot denotes a
time derivative and a dash a radial derivative.


\subsection{Polar-areal coordinates}

We introduce auxiliary variables $\Pi, P$ and $Y$ defined below by
\eqref{e:dtw}, \eqref{e:dtomega} and \eqref{e:udash} and write the YM
equations in first-order form (in time):
\begin{align}
  \label{e:dtw}
  \dot{w} &= A e^{-\delta}\,\Pi + u\,\omega,
  \\
  \label{e:dtomega}
  \dot{\omega} &= A e^{-\delta}\,P-u\,w,
  \\
  \label{e:dtPi}
  \dot{\Pi} &= \big(A e^{-\delta}\,w'\big)' + u\,P +
              w\,\frac{1-w^{2}-\omega^{2}}{r^{2}}e^{-\delta},
  \\
  \label{e:dtP}
  \dot{P} &= \big(A e^{-\delta}\,\omega'\big)' - u\,\Pi +
              \omega\,\frac{1-w^{2}-\omega^{2}}{r^{2}}e^{-\delta},
  \\
  \label{e:udash}
  \frac{r^{2}}{2}u' &= - Y e^{-\delta},\\
  \label{e:Ydash}
  Y' &= \omega\,\Pi - w\,P.
\end{align}
The Einstein equations and polar slicing condition reduce to
\begin{align}
  \label{e:dtA}
  \dot{A} &= 2r e^{-\delta}A^{3/2}J_{r},
  \\
  \label{e:Adash}
  A' &= \frac{1-A}{r} - 2r\rho,
  \\
  \label{e:deltadash}
  \delta' &= -\frac{r}{A}\left(\rho+S^{r}_{r}\right),
\end{align}
where the components of the energy-momentum tensor are
\begin{align}
  \label{eq:1}
  \rho &= \frac{Y^2}{r^4} + \frac{\left(1-w^2-\omega^2\right)^2}{4r^4}
  \\\nonumber
       & \qquad + \frac{A}{2r^{2}} \left(
                     P^{2} + \Pi^{2} + w'^{2} + \omega'^{2}\right),
  \\
  \label{eq:2}
  J_{r} &= - \frac{\sqrt{A}}{r^{2}}(\Pi w' + P \omega'),
  \\
  \label{eq:3}
  \rho + S^{r}_{r} &= \frac{A}{r^{2}} \left(
                     P^{2} + \Pi^{2} + w'^{2} + \omega'^{2}\right).
\end{align}
We fix residual gauge freedom taking coordinated $t$ to be proper time
of central observer, i.e. we set $\delta(t,r=0)=0$.

In the analysis of type II critical collapse we plot the two gauge
invariants
\begin{multline}
  \label{eq:12}
  I_{1} = -\frac{1}{8}\mathcal{F}^{(a)}_{\mu\nu}\mathcal{F}^{(a)\mu\nu} =
  \\
  \frac{Y^{2}}{r^{4}}
  - \frac{\left(1-w^{2}-\omega^{2}\right)^{2}}{4r^{4}}
  + A\frac{\Pi^{2} + P^{2} - w'^{2} - \omega'^{2}}{2r^{2}},
\end{multline}
\begin{multline}
  \label{eq:13}
  I_{2} = \frac{1}{8} \mathcal{F}^{(a)}_{\mu\nu}
  \left(\ast\,\mathcal{F}^{(a)\mu\nu}\right) =
  \\
  2 Y \frac{1-w^{2}-\omega^{2}}{r^{4}} +
  2 A \frac{Pw' - \Pi\omega'}{r^{2}},
\end{multline}
where the YM field strength tensor is
\begin{equation}
  \label{eq:14}
  \mathcal{F}^{(a)}_{\mu\nu} = \nabla_{\mu}\mathcal{A}^{(a)}_{\nu}
  - \nabla_{\nu}\mathcal{A}^{(a)}_{\mu} +
    \varepsilon^{abc} \mathcal{A}^{(b)}_{\mu} \mathcal{A}^{(c)}_{\nu}
\end{equation}
and its Hodge dual is
\begin{equation}
  \label{eq:16}
  \ast\mathcal{F}^{(a)}_{\mu\nu} = \sqrt{-g}\varepsilon_{\mu\nu\alpha\beta}
  \mathcal{F}^{(a)\alpha\beta}.
\end{equation}
The invariant $I_1$ is the Lagrangian of the YM field. The invariant
$I_2$ has the interesting property that it vanishes in the magnetic
sector.


\subsection{CMC-isotropic coordinates}

Following \cite{Rinne2013}, we introduce auxiliary variables
$D_F, D_H$ and $D_L$ defined below by \eqref{e:dtF}, \eqref{e:dtH} and
\eqref{e:Gdash} and write the YM equations in first-order form (in
time).
\begin{align}
  \label{e:dtF}
  \dot F &= rX F' - \tN D_F + 2XF - r^2 G H,\\
  \label{e:dtH}
  \dot H &= rXH' - \tN D_H + r^{-1} G' + GF + 3XH,\\
  \label{e:dtDF}
  \dot D_F &=(rX D_F - \tN F')' + 2X D_F -4 \tN r^{-1} F'\nonumber\\
    &-2 r^{-1} \tN' + G (D_L -r^2 D_H)\nonumber\\
    &+ \tN ( - 3F^2 - r^2H^2 + r^2F^3 + r^4FH^2),\\
  \label{e:dtDH}
  \dot D_H &= (rXD_H - \tN H')' - r^{-1}(XD_L)'\nonumber\\
    &-3 r^{-1} \tN'H + D_F(G - 2Xr^2H) \nonumber\\
    &+ XD_H(1 + 2r^2F) - 2XFD_L \nonumber\\
    &+\tN(-4r^{-1} H' + 2HrF' - 2FrH')\nonumber\\
    &+ \tN (-4FH + r^2 F^2 H + r^4 H^3 ),\\
  \label{e:Gdash}
  0 &= -\tN D_L + rG' + G,\\
  \label{e:DLdash}
  0 &= r^{-1} D_L' + 2F(D_L - r^2 D_H) + 2 D_H \nonumber\\
  &+ 2 r^2 H D_F.
\end{align}
We solve the following Einstein equations and coordinate conditions:
\begin{align}
  \label{e:cmc_hamcons}
  0 &= -4 \Omega \Omega'' + 6 \Omega'^2 - 8 \Omega r^{-1}\Omega'
  + \tfrac{3}{2} \Omega^2 r^4 \pi^2  \nonumber\\
  &- \tfrac{2}{3} K^2 + 2 \kappa \Omega^4 \tilde \rho, \\
  \label{e:cmc_momcons}
  0 &= \Omega (r\pi' + 5 \pi) - 2r\Omega' \pi
  + \kappa \Omega^3 r^{-1} \tilde J^r, \\
  \label{e:cmc_slicing}
  0 &= -\Omega^2 \tN'' + 3 \Omega\Omega'\tN' - 2\Omega^2 r^{-1}\tN' \nonumber\\
  &- \tfrac{3}{2}\Omega'^2 \tN + \sixth \tN K^2
  + \tfrac{15}{8} \tN \Omega^2 r^4 \pi^2\nonumber\\
  &+ \half \kappa \tN \Omega^4 (\tilde S + 2 \tilde \rho) , \\
  \label{e:isotropic}
  X' &= -\tfrac{3}{2} r \tN \pi.
\end{align}
Here $\pi$ denotes the only independent component of the traceless
part of the ADM momentum in spherical symmetry \cite{Rinne2013}. The
components $\tilde \rho$, $\tilde S$ and $\tilde J^r$ of the
(conformally rescaled) energy-momentum tensor are given by
\begin{align}
  \tilde \rho = \tilde S &= \half[3D_L^2 - 2r^2(2D_L D_H - D_F^2 - r^2 D_H^2)
     \nonumber\\ &+ 3 B_L^2 - 2r^2(2 B_L B_H - B_F^2 - r^2 B_H^2)],\\
  r^{-1} \tilde J^r &= 2[D_L B_F - D_F B_L \nonumber\\ &
     + r^2(D_F B_H - D_H B_F)], 
\end{align}
where we have defined the magnetic field components
\begin{align}
  B_F &= -3H - rH',\\
  B_H &= r^{-1} F' + r^2 H^2 + F^2,\\
  B_L &= -2F + r^4 H^2 + r^2F^2.
\end{align}


\section{Linear perturbations of static solutions}
\label{s:linpert}

In this section we write down the equations governing linear
perturbations of static EYM solutions explicitly and describe the
procedures used to solve the linearized system of equations.  We focus
on static solutions with a regular center, and as argued in
\cite{Galtsov1989,Ershov1990} we assume that the static solutions are
purely magnetic.

Assuming time independence, i.e. $w(t,r)=w_{s}(r)$,
$\delta(t,r)=\delta_{s}(r)$, $A(t,r)=A_{s}(r)$, and the magnetic
ansatz, i.e. $\omega=Y=u=0$, Eqs. \eqref{e:dtw}--\eqref{e:dtP} reduce
to
\begin{align}
  \label{e:static_eqns1}
  w_s'' &= \left(\frac{\left(w_s^2-1\right)^2}{2r^3A_s}
          +\frac{1-\frac{1}{A_s}}{r}\right)w_s'
          +\frac{w_s\left(w_s^2-1\right)}{r^2A_s},\\
  \label{e:static_eqns2}
  \delta_s' &= -\frac{w_s'^2}{r},\\
  \label{e:static_eqns3}
  A_s' &= \frac{1-A_s\left(w_s'^2+1\right)}{r}
         -\frac{\left(w_s^2-1\right)^2}{2 r^3}.
\end{align}
Regular solutions to \eqref{e:static_eqns1}--\eqref{e:static_eqns3}
are the Bartnik-McKinnon solitons $X_n$ \cite{Bartnik1988}. For the
purpose of the following analysis it is important to note the
asymptotic $r\rightarrow\infty$ expansion of the static solutions,
which reads
\begin{align}
  \label{e:static_asympt1}
  w_s(r) &= \pm 1 + \frac{v_{1}}{r} + \mathcal{O}\left(r^{-2}\right),
  \\
  \label{e:static_asympt2}
  A_s(r) &= 1 + \frac{a_{1}}{r} + \mathcal{O}\left(r^{-4}\right),
  \\
  \label{e:static_asympt3}
  \delta_s(r) &= \delta_{0} + \mathcal{O}\left(r^{-4}\right),
\end{align}
where the higher order terms are uniquely determined by the $v_{1}$,
$a_{1}$, and $\delta_{0}$.

Next, with a perturbative ansatz of the form ($|\ep|\ll 1$)
\begin{align}
  w(t,r) &= w_{s}(r) + \ep w_{p}(t,r), \\
  \label{e:pert_ansatz1}
  \omega(t,r) &= \ep \omega_{p}(t,r), \\
  \label{e:pert_ansatz2}
  u(t,r) &= \ep u_{p}(t,r), \\
  \label{e:pert_ansatz3}
  Y(t,r) &= \ep Y_{p}(t,r), \\
  \label{e:pert_ansatz4}
  A(t,r) &= A_{s}(r)\left(1 + \ep A_{p}(t,r)\right), \\
  \label{e:pert_ansatz5}
  \delta(t,r) &= \delta_{s}(r) + \ep \delta_{p}(t,r),
\end{align}
we obtain the following set of linearized equations:
\begin{align}
  \label{e:pert_eqns1}
  \frac{e^{2\delta_{s}}}{A_{s}}\ddot{w}_{p} &= A_s w_p'' + \left(\frac{1-A_s}{r}-\frac{\left(w_s^2-1\right)^2}{2 r^3}\right) w_p'
  \\ \nonumber
                                            & + \left(\frac{1-3 w_s^2}{r^2} - \frac{2 w_s \left(w_s^2-1\right)w_s'}{r^3}\right) w_p
  \\ \nonumber
                                            & + A_p \left(\frac{\left(w_s^2-1\right)^2 w_s'}{2 r^3} + \frac{w_s \left(w_s^2-1\right)}{r^2}-\frac{w_s'}{r}\right),
  \\
  \label{e:pert_eqns2}
  \delta_p' &= -\frac{2}{r} w_s' w_p',
  \\
  \label{e:pert_eqns3}
  A_p' &= -\frac{2 w_s \left(w_s^2-1\right) w_p}{r^3 A_s}
  \\ \nonumber
                                            & + \left(\frac{\left(w_s^2-1\right)^2}
                                              {2 r^2}-1\right)
                                              \frac{A_p}{r A_s}-\frac{2}{r} w_s' w_p',
  \\
  \label{e:pert_eqns4}
  \dot{A}_p &= -\frac{2}{r} w_s' \dot{w}_p,
  \\
  \label{e:pert_eqns5}
  \frac{e^{2\delta_{s}}}{A_{s}}\ddot{\omega}_{p} &= A_s \omega_p'' + \left(\frac{1-A_s}{r}-\frac{\left(w_s^2-1\right)^2}{2 r^3}\right) \omega_p'
  \\ \nonumber
                                            & - \frac{e^{2 \delta_s} w_s \dot{u}_p}{A_s}+\frac{\left(1-w_s^2\right) \omega_p}{r^2},
  \\
  \label{e:pert_eqns6}
  \dot{Y}_p &= A_s e^{-\delta_s} \left(w_s' \omega_p-w_s \omega_p'\right),
  \\
  \label{e:pert_eqns7}
  Y_p' &= -\frac{e^{\delta_s} w_s \left(w_s u_p+\dot{\omega}_p\right)}{A_s},
  \\
  \label{e:pert_eqns8}
  u_p' &= -\frac{2}{r^2} e^{-\delta_s} Y_p.
\end{align}
To simplify (\ref{e:pert_eqns1})--(\ref{e:pert_eqns8}) we used the
equations \eqref{e:static_eqns1}--\eqref{e:static_eqns3} satisfied by
static solutions.  This explicitly shows that the linear perturbation
splits into two independent classes: magnetic sector
\eqref{e:pert_eqns1}--\eqref{e:pert_eqns4} and sphaleronic sector
\eqref{e:pert_eqns5}--\eqref{e:pert_eqns8}. We analyze them
individually below.


\subsection{Magnetic perturbations}

Separation of variables
\begin{equation}
  \label{e:magnetic_sep}
  w_{p}(t,r) = \phi(r)e^{i\sigma t}, \ A_{p}(t,r) = \alpha(r)e^{i\sigma t}, \
  \delta_{p}(t,r) = \beta(r)e^{i\sigma t}
\end{equation}
reduces \eqref{e:pert_eqns1}--\eqref{e:pert_eqns4} to a system of
ordinary differential equations
\begin{align}
  \label{e:magnetic_pert1}
  \frac{e^{2\delta_{s}}\sigma^{2}}{A_{s}^{2}}\phi &=
  -\phi'' + \left(-(1-A_{s}) + \frac{(1-w_{s}^{2})^{2}}{2r^{2}}\right)\frac{1}{rA_{s}}\phi'
  \nonumber\\
  & \quad + \Big(-\frac{4w_{s}(1-w_{s}^{2})w_{s}'}{r}
    + \frac{(1-w_{s}^{2})^{2}w_{s}'^{2}}{r^{2}}
   \nonumber\\
  & \qquad
    + (-1+3w_{s}^{2}-2w_{s}'^{2})
    \Big)\frac{1}{r^{2}A_{s}}\phi,
  \\
  \label{e:magnetic_pert2}
  \alpha &= -\frac{2}{r}\phi w_s',
  \\
  \label{e:magnetic_pert3}
  \beta' &= -\frac{2}{r}w_s'\phi'.
\end{align}
Note that \eqref{e:magnetic_pert1} does not contain any metric
perturbations; therefore the solution to \eqref{e:magnetic_pert1}
fully determines the perturbation \eqref{e:magnetic_sep} through the
relations \eqref{e:magnetic_pert2}--\eqref{e:magnetic_pert3}.

\subsubsection{Unstable modes}

Using standard methods (either shooting or a pseudospectral method) we
look for solutions of \eqref{e:pert_eqns1}--\eqref{e:pert_eqns4}
imposing asymptotically flat boundary conditions at spatial infinity.
We find the value of the exponent of the unstable mode of $X_{1}$ to
be $\lambda = i\sigma = 2.562799802146866$. We also find, in agreement
with previous studies \cite{Lavrelashvili1995}, $n$ unstable modes of
the solution $X_{n}$. (We do not explicitly give the values for
$X_{n>1}$ as these have more than one unstable mode and thus do not
play any role in the critical collapse evolutions we consider here.)

\subsubsection{Quasinormal modes}
\label{sec:quasi-normal-modes}

To find QNM we use the same shooting method as when looking for
unstable modes. However, we now impose an outgoing boundary condition
at spatial infinity. Taking
\begin{equation}
  \label{eq:4}
  \phi(r) = e^{-irc_{\infty}}\xi(r), \quad c_{\infty}=e^{\delta_{0}},
\end{equation}
where $\delta_{0}$ is the asymptotic value of $\delta(r)$
[cf. (\ref{e:static_asympt3})] and changing the independent variable
to $z=1/r$ we transform Eq.~(\ref{e:magnetic_pert1}) to
\begin{equation}
  \label{eq:5}
  \xi''(z) + P(z)\xi'(z) + Q(z)\xi(z) = 0.
\end{equation}
The coefficients in the above equation (determined by the static
solution and $\sigma$) have the following asymptotic form as
$z \to 0$:
\begin{equation}
  \label{eq:7}
  P(z) = \frac{p_{-2}}{z^{2}} + \mathcal{O}\left(z^{-1}\right),
  \quad
  Q(z) = \frac{q_{-3}}{z^{3}} + \mathcal{O}\left(z^{-2}\right),
\end{equation}
with the expansion coefficients depending on $v_{1}$, $a_{1}$,
$\delta_{0}$, and $\sigma$. Thus $z=0$ is an irregular singular point
of Eq.~(\ref{eq:5}).  However, assuming
\begin{equation}
  \label{eq:6}
  \xi(z) = z^{k}\sum_{i\geq 0}\xi_{i}z^{i},
\end{equation}
the indicial equation gives
$k=-q_{-3}/p_{-2}=-i e^{\delta_{0}} a_{1}\sigma$, and we uniquely
determine the expansion coefficients $\xi_{i}$ [which are given in
terms of the asymptotic expansion
(\ref{e:static_asympt1})-(\ref{e:static_asympt3})].

Having two asymptotic solutions, one at the origin and the other
obtained from the above asymptotic analysis, we integrate the
Eq.~(\ref{e:static_asympt3}) starting from the two boundary points.
Gluing the solutions at an intermediate point gives a quantization
condition for $\sigma$. With this procedure we find the least damped
QNM of $X_{1}$, whose frequency is $\lambda = -1.40233 \pm 3.60351i$.
Interestingly enough with this method we were also able to obtain
higher overtones (with faster damping rates) but these were not
independently confirmed by time evolution and so we omit their
presentation here.


\subsection{Sphaleronic perturbations}

Separation of variables
\begin{equation}
  \label{e:sphaleron_sep}
  \omega_{p}(t,r) = \psi(r)e^{i\sigma t}, \ Y_{p}(t,r) = y(r)e^{i\sigma t},
  \ u_{p}(t,r) = \upsilon(r)e^{i\sigma t},
\end{equation}
reduces \eqref{e:pert_eqns1}--\eqref{e:pert_eqns4} to
\begin{align}
  \label{e:sphaleron_pert1}
  -\frac{e^{2\delta_s}\sigma^2}{A_s}\psi &= A_s \psi '' + \left(\frac{1-A_s}{r}-\frac{\left(w_s^2-1\right)^2}{2r^3}\right)\psi'
  \\ \nonumber
  & \quad  -\frac{i \sigma \upsilon e^{2\delta_s} w_s}{A_s}  +\frac{\psi
   \left(1-w_s^2\right)}{r^2} ,
  \\
  \label{e:sphaleron_pert2}
  y &= \frac{i A_s e^{-\delta_s} \left(w_s \psi'-\psi w_s'\right)}{\sigma},
  \\
  \label{e:sphaleron_pert3}
  \upsilon' &= \frac{2 i A_s e^{-2 \delta_s} \left(\psi w_s'-w_s \psi'\right)}{r^2 \sigma}.
\end{align}

\subsubsection{Unstable modes}

In this sector we also find (as for the magnetic ansatz) $n$ unstable
modes for $X_{n}$. For the fundamental solution $X_1$ we have
$\lambda = i\sigma = 2.7831012067733285$.

\subsubsection{Quasinormal modes}

In the nonlinear evolution we see no sign of quasinormal modes within
the sphaleronic sector. Thus we leave open the question of their
existence.


\section{Numerical methods}
\label{s:nummethods}

In this section we briefly describe the numerical methods used in our
two independent codes.


\subsection{Type I and type II collapse}

For the time evolution we use the method of lines with a second-order
finite-difference discretization in space and the explicit Runge-Kutta
time integration scheme DOPRI (a fifth-order adaptive method)
\cite{Hairer1993}. To refine the central region of the spatial domain
we use a nonequidistant grid.  The spacing between grid points is
fixed over time.  We choose a logarithmic distribution which
concentrates grid points close to $r=0$ and has physical extent
$r\in [0, r_{m}]$, explicitly
\begin{equation}
  \label{e:rmapping}
  r_{i} = r_{m} \log\left(1-\left(\frac{i}{N}\right)^{k}\right)/
  \log\left(1-\left(\frac{N-1}{N}\right)^{k}\right),
\end{equation}
$i=0, 1, \ldots N-1$. The two free parameters $k$ and $r_{m}$ in
(\ref{e:rmapping}) were chosen to reach a compromise between higher
resolution close to the origin (sufficient to represent fine
structures of solutions) and a sufficiently large physical extent of
the grid (so that the numerical solution is not affected by the
presence of a timelike boundary).  At the outer boundary we use
one-sided finite-difference stencils.  Most of the simulations were
carried out using $k=3/2$ and $r_{m}=200$ or $r_{m}=400$.  We
typically take from $N=1+2^{10}$ to $N=1+2^{12}$ grid points.


\subsection{Type III collapse}

This code uses the method of lines with a fourth-order
finite-difference discretization in space and the standard
fourth-order Runge-Kutta method for the time evolution.  Ordinary
differential equations with respect to radius are solved using a
Newton-Raphson method combined with a direct band-diagonal solver. In
the first phase of the evolution, the radial grid is uniform and
ranges from the origin to future null infinity, where one-sided finite
differences are used. When a black hole forms, an excision boundary is
placed just inside the apparent horizon, where again one-sided
stencils are used. The YM variable $G$ is fixed to zero at the
excision boundary. In this second phase of the evolution, the radial
grid is nonuniform in order to provide more resolution close to the
horizon, where the fields have large gradients.  Typical resolutions
range from $500$ (Figs. \ref{f:type3_w_wtilde}--\ref{f:type3_mag_dtw})
to $4000$ (Figs. \ref{f:type3_sph_dtw} and \ref{f:type3_sph_dtomega})
radial grid points.  More details on the numerical implementation can
be found in \cite{Rinne2013,Rinne2014a}.
\\



\input{paper.bbl} 

\end{document}

%% file: paper.bbl
%